\documentclass[twoside,a4,12p]{report} %,draft,openright]
\usepackage{color}
\usepackage{epsf,graphicx}
\usepackage{latexsym,amssymb}
\usepackage{setspace,cite}
\usepackage{amsmath}
 \usepackage{subfig}
\usepackage{anysize}
\marginsize{4cm}{2.5cm}{4cm}{4cm}
\usepackage{multirow}
\usepackage{fancyhdr}

\begin{document}
\newpage

%Puts page numbering of preamble in roman and of main body of thesis in
%arabic. Also defines how chapters and sections are made
\pagenumbering{arabic}
\setcounter{page}{1} \pagestyle{fancy}
\renewcommand{\chaptermark}[1]{\markboth{\chaptername%
\ \thechapter:\,\ #1}{}}
\renewcommand{\sectionmark}[1]{\markright{\thesection\,\ #1}}

%DEFINES TITLE PAGE, and contains abstract, acknowledgements, etc.

%%%%%%%%%%%%%%%%%%%%%%%%%%%%%%%%%%%%%%%%%%%%%%%%%%%%%%%%%%%%%%%%%%%%%%%%%%%
% This is a sample header for a sample dissertation. Fill in the name,
% and the other information. LaTeX will work out the table of
% content, the list of figures and of tables for you.
%%%%%%%%%%%%%%%%%%%%%%%%%%%%%%%%%%%%%%%%%%%%%%%%%%%%%%%%%%%%%%%%%%%%%%%%%%%

\newpage
\thispagestyle{empty}

% ******* Title page *******
% **************************

\vspace*{2cm}
\begin{center}
{\Large\bf Freehand 2D Ultrasound Probe Calibration for Image Fusion with 3D MRI/CT \\} \vspace{2cm} {\large 
Yogesh Langhe\\
\vspace{2cm}
%Universit\'{e} de Bourgogne \\Universitat de Girona \\ Heriot-Watt University
}

\end{center}
\begin{center}
{\large \textbf{Supervised By} \\ Dr. Katrin Skerl and Prof. Adrien Bartoli \\ EnCoV-ISIT \\ UMR 6284 CNRS/Universit\'{e} d'Auvergne}
\end{center}

\vspace{3cm}
\begin{center}
{\large A Thesis Submitted for the Degree of \\MSc Erasmus Mundus
in Vision and Robotics (VIBOT) \\\vspace{0.3cm} $\cdot$ 2017
$\cdot$}
\end{center}
\begin{figure}[!htb]
	\centering
    \includegraphics[width=1.7cm,height=1.7cm]{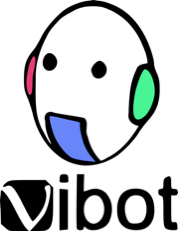} \quad
	\includegraphics[width=2cm,height=1.7cm]{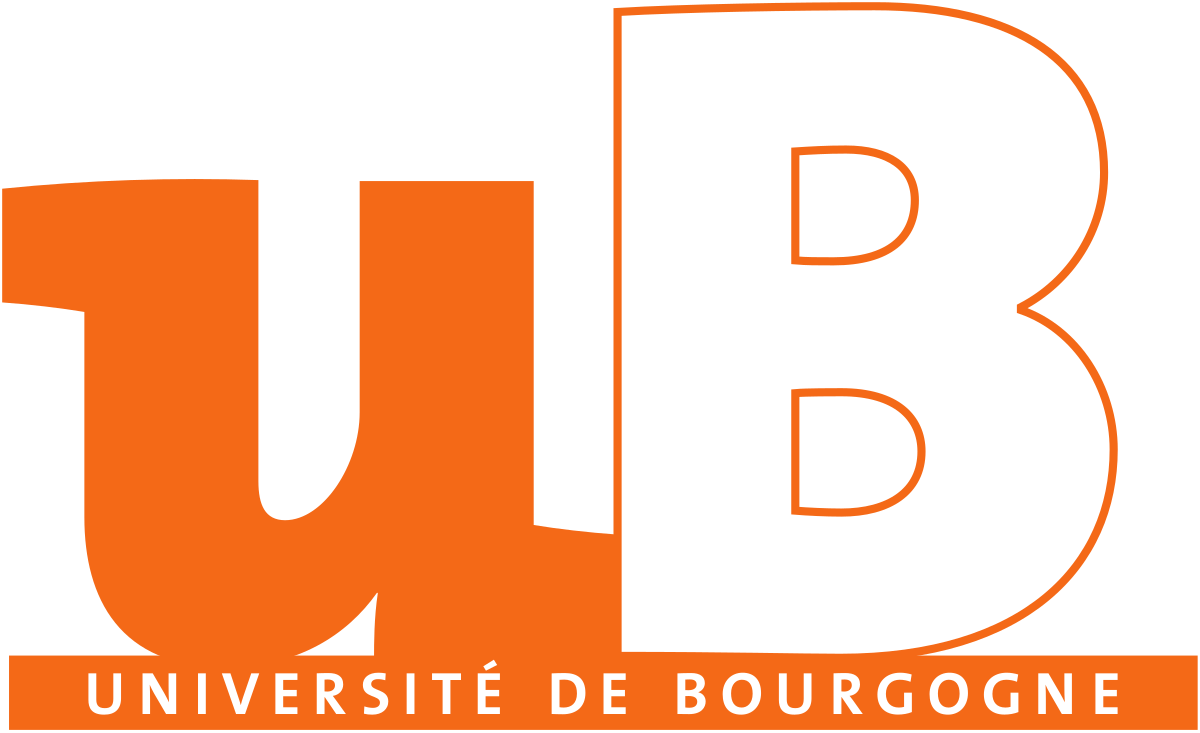} \quad
	\includegraphics[width=2cm,height=1.7cm]{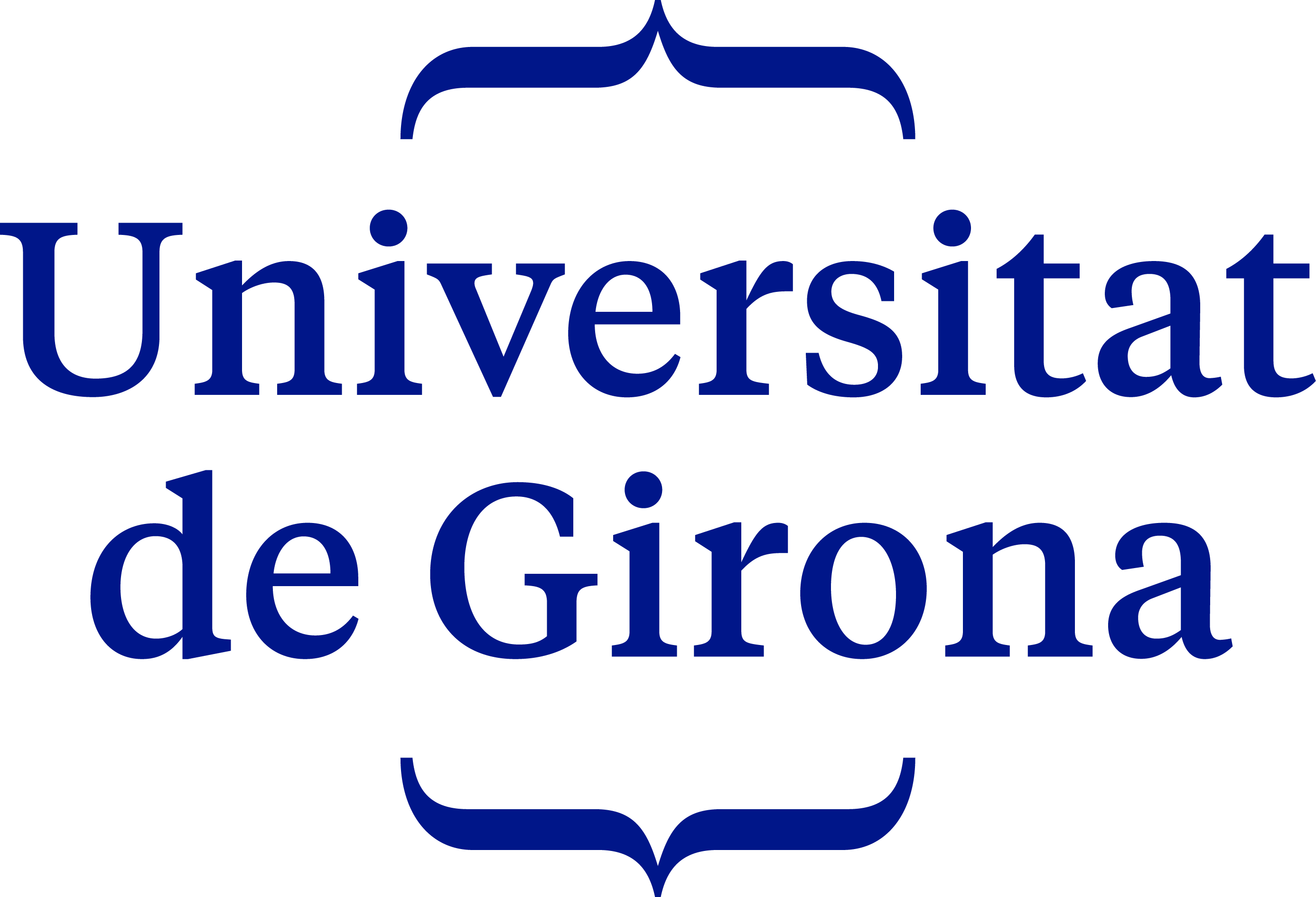}\quad
	\includegraphics[width=2cm,height=1.7cm]{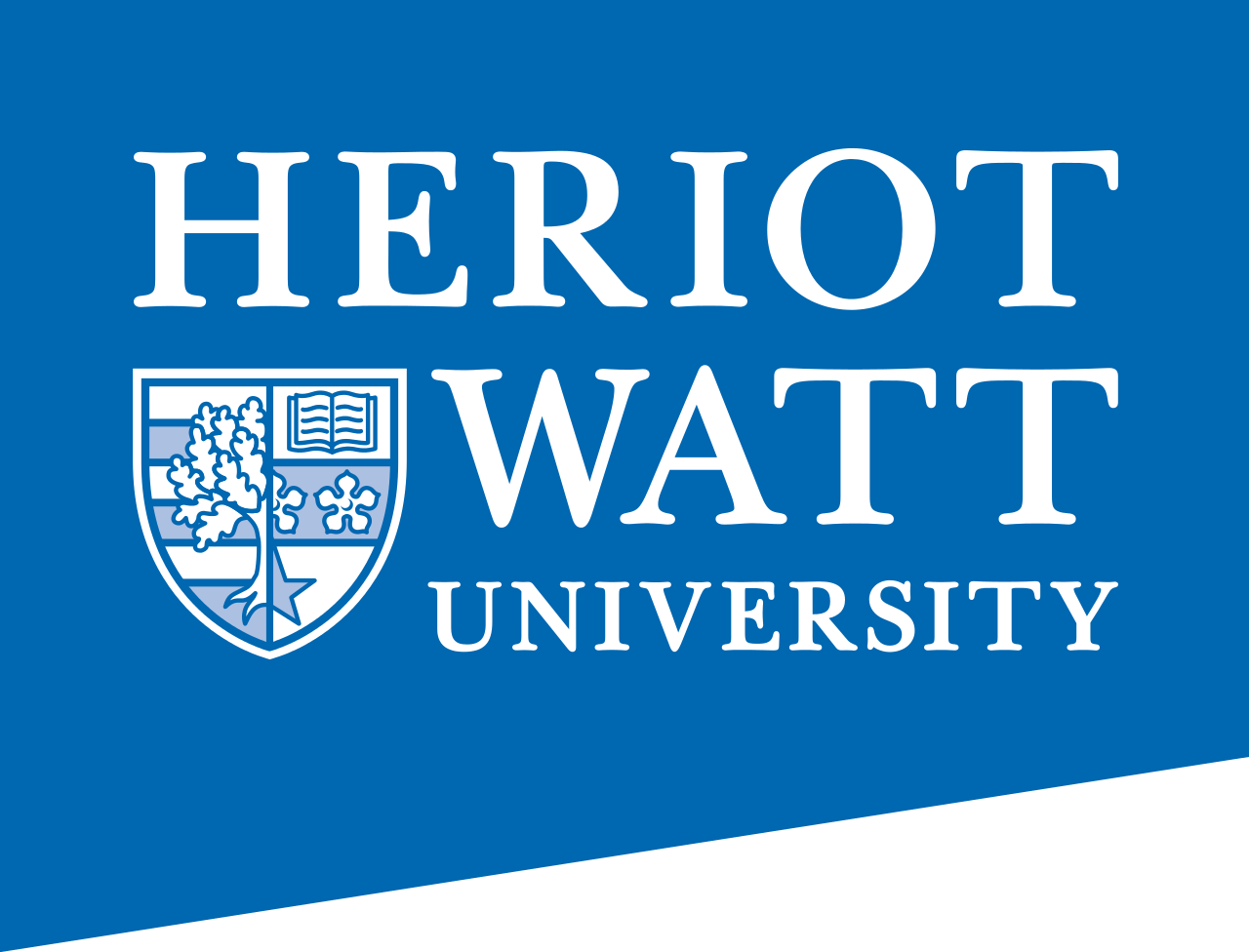}\quad
	\includegraphics[width=2cm,height=1cm]{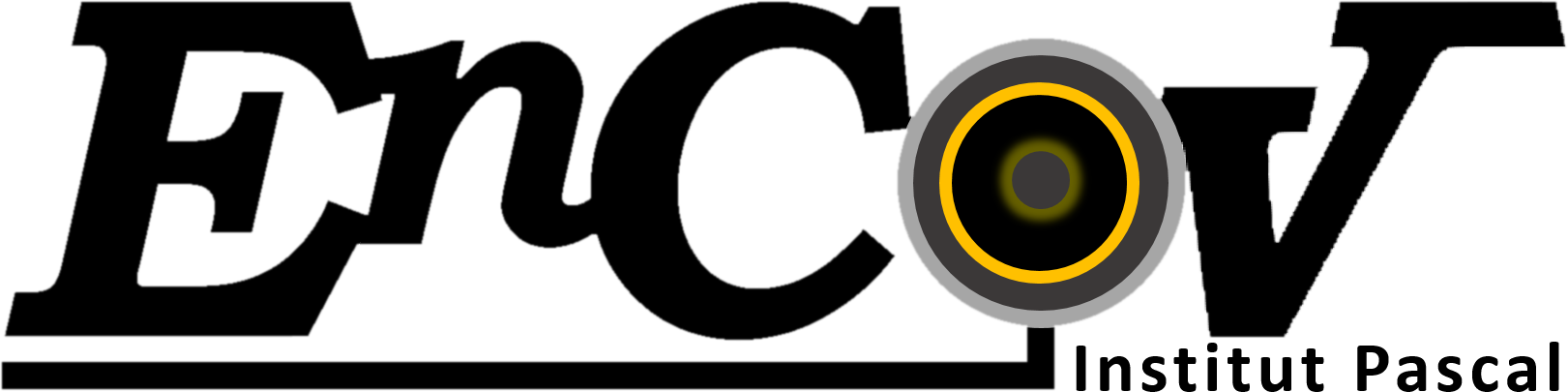}
\end{figure}
\singlespacing

\vspace{-2cm}
%ABSTRACT
\begin{abstract}
The aim of this work is to implement a simple freehand ultrasound (US) probe calibration technique.  The probe calibration problem estimates the 2D US image plane with respect to a fixed external reference system. This will enable us to visualize US image data during surgical procedures using augmented reality. Furthermore, it allows the application of image fusion with MRI/CT. The developed system is a preliminary step for the prospective image fusion application to improve its accuracy and speed. The calibration system is realized by tracking the US probe equipped with a position sensor i.e it aims to estimate the spatial transformation between position sensor coordinate system and US image coordinate system. The calibration procedure estimates a rigid transformation to map all US B-mode image pixels into the pose tracking sensor co-ordinate system.  The problem of estimating the transformation between two coordinate systems is referred as absolute orientations. It is solved by using least-squares minimization. This transformation is estimated using the features in the physical space and its correspondences in the US images. An object with predefined geometry (known as calibration phantom) is designed to obtain these sets of features and simultaneously localizing them in the US images. The accuracy of the system is determined by the ability to localize these features in the physical space and US images. Three different calibration phantoms were designed and tested for the procedure. 3D printed calibration phantoms were used in this process. However, only one phantom was used to realize the calibration system. The pose estimation with 6-degrees of freedom was performed using a single camera based optical system and a fiducial markers rig.
The performance of the system was evaluated with different experiments using two different pose estimation techniques. A near-millimeter accuracy can be achieved with the proposed approach. The developed system is cost-effective, simple and rapid with low calibration error. 

\vspace*{5cm}

\begin{center}
\begin{quote}
\it Research is what I'm doing when I don't know what I'm
doing.\,\ldots
\end{quote}
\end{center}
\hfill{\small Werner von Braun}

\end{abstract}

\doublespacing

\pagenumbering{roman}
\setcounter{page}{1} \pagestyle{plain}

\tableofcontents
\listoffigures
\listoftables

\chapter*{Acknowledgments}
\addcontentsline{toc}{chapter}
         {\protect\numberline{Acknowledgments\hspace{-96pt}}}

I express my sincere gratitude to my supervisors Dr. Katrin Skerl and Prof. Adrien Bartoli for their support, guidance and a rewarding learning experience during this internship.
I would specially like to thank Dr. Lilian Calvet for his expertise and help throughout this work. In addition, I would like thank all the other members and friends at EnCoV group for the co-operation. 
I would also like to thank my family and friends for all the love and support. 
Finally, I am grateful to European Union for the generous financial support during this masters program. 

\pagestyle{fancy}

\newpage

%sets up headers for lefthand and righthand pages. To alter, edit
%these lines and the chaptermark/sectionmark lines above
\addtolength{\headheight}{3pt} \fancyhead{}
\fancyhead[LE]{\sl\leftmark} \fancyhead[LO,RE]{\rm\thepage}
\fancyhead[RO]{\sl\rightmark} \fancyfoot[C,L,E]{}
\pagenumbering{arabic}

%\singlespacing
%\doublespacing
\onehalfspacing
\chapter{Introduction} \label{chap:intro}

\section{Motivation} \label{sect:MotivationIntro}
In the 20th century, Minimal Invasive Surgery (MIS) was introduced in order to reduce the average recovery time of the patient after the surgery. There the surgical procedure is performed via small incision for the surgical tools and a camera. However, the surgeons lose the ability of direct contact with the tissue. Furthermore, the depth perception is lost. Thus, the computer aided assistance systems were developed. One of such systems is Augmented Reality (AR) which assists surgeons by providing a virtual transparency of the patient in real time. 

AR in computer aided surgeries enables surgeons to visualize image data such as obtained pre-operatively by overlaying the real time video stream captured with the endoscopic camera. One such application field is laparoscopic surgery, i.e. laparosurgery, of the uterus. AR supports laparosurgery by showing the organ’s anatomical structures in 3D. The Endoscopy and Computer Vision (EnCoV) group at the Universit\'{e} d'Auvergne developed an algorithm allowing real-time fusion of the laparoscopic video stream with pre-operative Magnet Resonance Images (MRI). It enables to display structures such as lesions and guide for the surgeons to place the incisions using Shape from Template (SfT). However, this model is only applicable to more rigid organs such as the uterus. To extend the application to other less rigid organ sites such as the liver a fusion with elastography imaging is beneficial. The stiffness of the organ can be measured with Elastography Imaging. It is a novel imaging technique visualizing the tissue’s elasticity. Moreover, it supports staging of liver fibrosis and cancer diagnosis \cite{bamber2013efsumb}. Some MRI/CT based elastography technique are available in recent years but the most common and cost-effective elastography technique is based on US imaging.

 The image registration of US based elastography images with MRI/CT is a difficult task due to intrinsic differences in gray-level intensities characteristics, the presence of artifacts including deformation and noise \cite{yavariabdi2013mapping}. The image registration accuracy and the speed is improved by estimating the US imaging plane within the tracking system.  This process of estimation of US image in 3D space is known as probe calibration (also referred as US calibration). This procedure enhances the mutual information metric and improves the accuracy of US-MRI/CT registration.

\section{Background}

US imaging is one of the vital imaging modality used medicine. It is widely used for its safety, cost-effectiveness and portability. Moreover it works in real time and does not require any special operating conditions. 
Conventional US imaging which provides 2D images has limited ability in the visualization of the three dimensional scanned anatomy. Thus, there has been active research in introducing 3D US that helps more complex diagnostics and provide better visualization. Although 3D US transducers (3D probe) which provide 3D images have been developed, their use is limited due to the artifacts, low resolution, higher cost, large probe size and short scanning area \cite{hung20073d}. Thereby, 2D probes are mostly used to reconstruct 3D volume. Such systems are realized mainly using three techniques \cite{hsu2009freehand}:
\begin{itemize}
\item Mechanically swept probes: A robotic arm or motors are used to mechanically sweep the probe in such systems. All 2D B-scans are subsequently combined to build a 3D volume. These techniques are useful while scanning small regions. However, they provide a short field of view \cite{hsu2009freehand}. 
 
 \item Sensorless freehand techniques:  In this approach, information from B-scans itself is used to build the volume. The separation of pair of frames is estimated using speckle decorrelation to build a 3D volume without a sensor\cite{housden2007sensorless}. The accuracy of such systems is poor compared to other techniques \cite{hsu2009freehand}.
 \item Freehand techniques: In these techniques, the probe is tracked mainly using tracking systems such as electromagnetic sensors or optical sensors. The 3D volume is reconstructed using non-uniformly spaced B-scans \cite{hsu2009freehand}.
 
\end{itemize}

In this work, we mainly focus on freehand techniques. When it is required to locate the volume in a fixed external coordinate system, freehand techniques are preferred. Furthermore, this provides information about the size and orientation of the patient or surgical bed while image registration with other modalities. 
In freehand techniques, a position sensor is equipped with the probe to track its position relative to the fixed reference system. This allows reconstruction of 3D volume. However, the rigid transformation (rotation + translation) between the tracking sensor and US image is unknown in such systems. The procedure of estimating this transformation is known as probe calibration or ultrasound calibration. Probe calibration and pose tracking of US probe enhance applications for augmented reality, 3D US volume reconstruction, image registration etc.

\section{Objective}
This project aims to solve a probe calibration problem in order to facilitate the US-MRI/CT image fusion. 
The main objective of this work is to design and develop a simple, rapid and cost-effective US probe calibration system. In the work presented, we design and evaluate different calibration phantoms. We also provide a cost-effective tracking system for the pose estimation with 6-DoF with respect to the fiducial markers.

\section{Thesis Outline}
\begin{itemize}
\item Chapter~\ref{chap:problemstatement} summarizes the probe calibration problem in brief.
\item Chapter~\ref{chap:StateofArt} provides the overview of the existing calibration methods related to our work. It also entails different types of calibration phantoms.
\item Chapter~\ref{chap:Methodology} discusses the details of the proposed system. This chapter describes the overall approach to the calibration problem and also explains the experimental procedure. 
\item Chapter~\ref{chap:implementation} outlines the details of the pose estimation and the phantom landmark detection method.
\item Chapter~\ref{chap:ExpAndResults} presents the experimental results of different phantoms and the calibration procedure.
\item Chapter~\ref{chap:Conclusion} concludes summary of the proposed system. In addition, the future extension to the work is discussed in this chapter.

\end{itemize}

%\subsection{US Image Registration}

%\subsection{3D Ultrasound Imaging}

%\section{Technical background}
 
%\subsection{Probe Calibration}
\chapter{Problem Statement} \label{chap:problemstatement}
\begin{figure}[!htb]
	\centering
	\includegraphics[width=7.3cm,height=7.3cm]{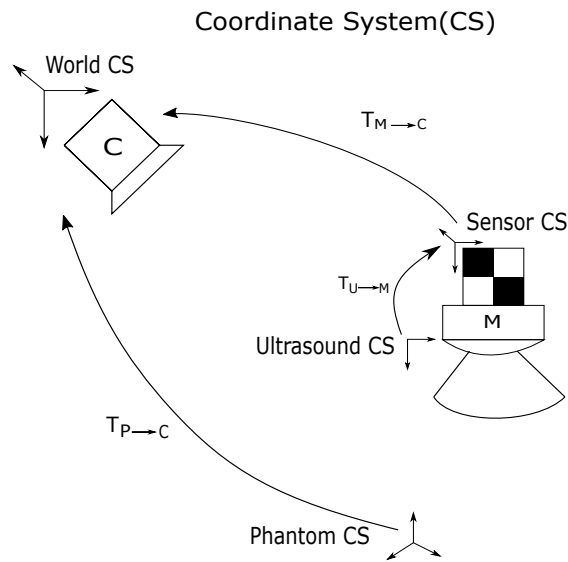}
	\caption{The associated coordinate systems }
\label{fig:problemstatement}
\end{figure}

This thesis aims to estimate the relative location of the B-scan with the position sensor attached to the probe. In other words, it aims to estimate the spatial relationship between the US image plane and the camera in a freehand US system. The role of calibration is to find the mathematical transformation that converts the 2D coordinates of pixels in the US image into 3D coordinates of the frame of reference (or world coordinate system). The rigid transformation $T_{U \rightarrow M}$ with 6 degrees of freedom (3 rotation + 3 translation) between the US imaging plane and the tracking sensor should be estimated (Figure~\ref{fig:problemstatement}).
 This transformation is determined by probe calibration. The probe calibration problem should be solved by estimating the features in the physical space $P$ and its correspondences in US B-mode images $U$ using Eq~\ref{eq:main_pro_state}. 
 \begin{equation} 
 P
 = T_{P\rightarrow C} T_{M\rightarrow C}  T_{U\rightarrow M}  U \tag{1}
 \label{eq:main_pro_state}
 \end{equation}
 
  These features are obtained by scanning a object with predefined geometry. If the scaling of the US image is known, all points are mapped into the world coordinate system (Figure~\ref{fig:problemstatement}) using Eq.~\ref{eq:main_pro_state}.  
The transformations $T_{P \rightarrow C}$ and $T_{M \rightarrow C}$ are estimated using the camera. 
This thesis aims to provide a reliable, easy and accurate probe calibration technique for the application to image fusion. 

\chapter{State of the art} \label{chap:StateofArt}

\section{Introduction}

In the last few decades, there is an active ongoing research to reconstruct 3D US volume from conventional 2D images. Although 3D transducers are available, their use is limited due to artifacts, short scanning area and poor quality \cite{hung20073d,hsu2009freehand}. The system of volumetric reconstruction of US is realized by placing a position sensor to the probe that tracks its pose in 3D space. Several methods have been proposed to construct freehand 3D US volumes. A comprehensive review of such methods and different calibration phantoms is discussed in Mercier et al. \cite{mercier2005review} and Hsu et al. \cite{hsu2009freehand}. 

Typically the proposed methods follow a similar procedure for the  calibration task. It involves tracking a probe with a position sensor to estimate the pose of the probe and scanning a calibration phantom. 
The overall calibration setup depends on the objectives of the procedure \cite{hsu2009freehand}. The objectives can be either to optimize the calibration procedure or whether the features should be extracted manually or automatically. Furthermore, the aimed application such as volume reconstruction, image registration or needle guided therapy can determine the objectives. Most techniques involve a calibration phantom designed to provide features in B-scans and the physical space. These features are used to determine the correspondences for the calibration solution. Most methods differ in terms of the calibration phantoms used.  
\section{Calibration Phantoms}
A calibration phantom is a scanning object with predefined geometric properties. It is designed to obtain a precise calibration by localizing a set of features accurately in the B-mode US images (referred as B-scan here onwards). These features are also localized in the phantom space or phantom CS. Each calibration phantom provides different accuracy, speed of calibration and different acquisition methods (manual or automatic feature extraction)\cite{bakulina2016automated}. 
 The calibration phantoms are predominantly categorized in four classes
\cite{mercier2005review} described as follows:
%\subsection{Classes of phantoms}
\subsection{Single point phantoms}
\begin{figure}[htbp]
\centering
%\subfloat[]{
\includegraphics[width=4cm,height=4cm]{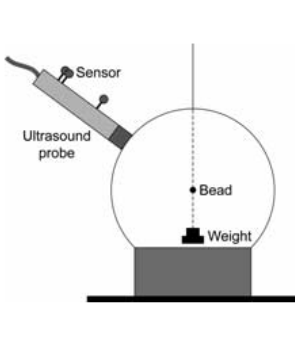} % }
%\quad
%\subfloat[]{
%\includegraphics[width=8cm,height=4cm]{figures/Crosswire.png}
%}
\caption{A possible realization of point phantom  \cite{mercier2005review}.}
\label{fig:point}
\end{figure}

Point phantoms (Figure~\ref{fig:point}) are small spherical objects like a pin head or a bead which can be scanned with US. Such phantoms are easy to construct but difficult to locate in the B-scan. The center of the phantom is segmented manually. Although automatic techniques to overcome this challenge, they may lead to many false positive results \cite{mercier2005review}. Point phantoms provide only a single point, thus it should be scanned many times for the calibration. There are other variations of the point phantoms like cross-wire \cite{detmer19943d,trobaugh1994three} phantoms. Cross-wire phantoms are  composed of two wires intersecting in a point. 
Although a distinct spot of the phantom is expected in each B-scan, the small size of the phantom makes it difficult to visualize in the B-scan. Thus, the appearance of this point can vary due to noise or artifacts in each B-scan. The accuracy of such phantoms rely on the ability to localize the phantom feature can be localized in the B-scan as well as in the phantom. 

\begin{figure}[htbp]
\centering
\includegraphics[width=8cm,height=4cm]{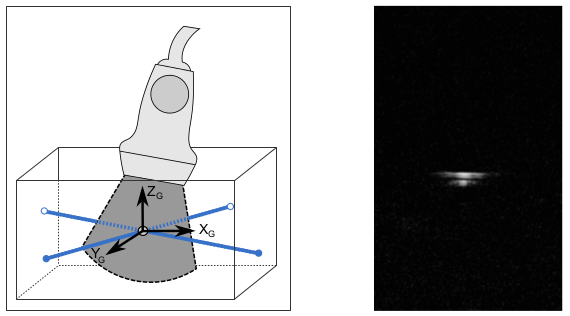}
\caption{Cross-wire phantom and its B-scan on the right \cite{bakulina2016automated}.} 
\label{fig:wire}
\end{figure}

\subsection{Multiple point targets and cross wire phantoms}
These phantoms are derived from point and cross-wire phantoms as discussed before. Multiple small objects \cite{lindseth2003probe,kowal2003development} or cross-wires\cite{trobaugh1994three} are scanned in this approach. An example of a single point cross-wire phantoms is shown in Figure~\ref{fig:wire}. Multiple cross-wires are used to realize multi cross-wire phantoms. The B-scans of such phantoms provide more number of feature points and/or lines with fewer scans compared to single point phantom. The drawback of these phantoms include alignment of the US plane with the multiple targets for simultaneously scanning\cite{bo2015versatile}.

\subsection{Wall phantoms and 2D shape alignment phantoms}
\begin{figure}[htbp]
	\centering
	\includegraphics[width=7cm,height=3cm]{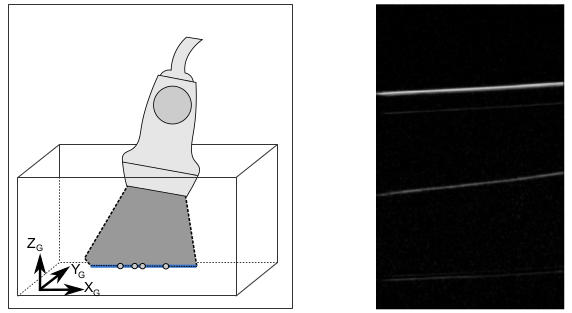}
	\caption{Wall phantom and its B-scan\cite{bakulina2016automated}.} 
	\label{fig:wallphantom}
\end{figure}

\begin{figure}[htbp]
\centering
\includegraphics[width=4cm,height=4cm]{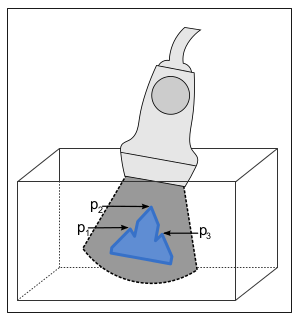}
\caption{Alignment phantom by Sato et al.\cite{sato1998image,bakulina2016automated}.} 
\label{fig:alignphantom}
\end{figure}

The idea behind such phantoms is to scan a plane surface or membranes which produce line in the B-scan. The line feature can be easily segmented compared to the point feature. Some authors have imaged the bottom of the water tank which produces lines in B-scan. The proper alignment of the US plane with a wall (Figure~\ref{fig:wallphantom}) or a membrane (Figure~\ref{fig:alignphantom}) is difficulty in these phantoms. The appearance of the lines depends on the angle of the probe relative to the water surface. There could be reverberation artifacts caused by multiple reflections in such phantoms\cite{bo2015versatile}.

\subsection{Z-fiducial phantoms}
\begin{figure}[htbp]
\centering
\includegraphics[width=8cm,height=4cm]{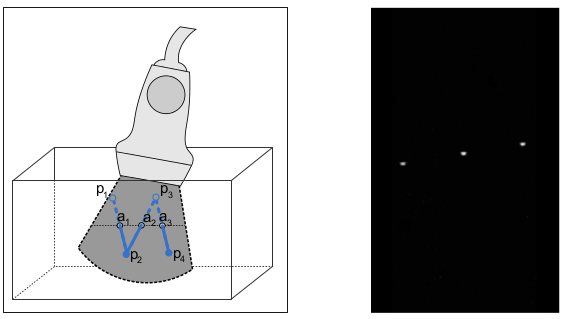}
\caption{Z-phantom and its B-scan on the right \cite{bakulina2016automated}.} 
\label{fig:zphantom}
\end{figure}

In these phantoms, the wires are arranged in between walls to form a Z-shape (Figure~\ref{fig:zphantom}). The B-scan of such a phantom shows points corresponding to each wire. The position of the end-points of the wires are and the relative distance between the points is known in the phantom space. These correspondences are used to perform the calibration procedure. With these phantoms, a large number of features can be measured with a single acquisition thereby reducing calibration time.

\section{Probe tracking}
Different types of tracking technologies have been exploited to track the probe, most commonly electromagnetic and optical tracking. Many methods use commercial electromagnetic sensors such as Flock of Birds (Ascension Technology Corporation,
U.S.A) and Aurora (Northern Digital) \cite{hsu2009freehand}. One of the major drawback with the electromagnetic system is that the receiver has to be placed near the transmitter within the generated magnetic field. Moreover, the accuracy is impeded by the metallic (AC sensors) and ferromagnetic material (DC sensors) \cite{hsu2009freehand}. Thus, such systems are prone to environmental noise and distortion due to metal objects in and near the tracking field. It is a challenging task to keep metallic objects (surgical instruments) far away from the tracking system in operating rooms \cite{mercier2005review}. 

In optical tracking, the system is realized with multiple cameras tracking the markers attached to the target (probe).
Both passive (retro-reflective material) or active markers (Infrared LEDs) are used for the pose estimation. There is not much noticeable difference between active and passive markers \cite{wiles2004accuracy}. Multiple markers are rigidly attached to the targets which needs to be tracked.  Optical tracking systems suffer from several other drawbacks, including a limited field of view, the need for a clear line of sight, and a lack of orientation accuracy with small markers \cite{mercier2005review}. Many research groups use commercial optical tracking systems such as Polaris from NDI (Waterloo, ON, Canada) and Optitrack systems to develop freehand US applications.  
 
 \section{Sensorless technique}
  These techniques do not require any tracking sensors to estimate the pose of the probe in the space. The pose is estimated  using the information from the US images itself. Some techniques \cite{housden2007sensorless,laporte2011learning} analyze the speckle in US images using decorrelation or linear regression methods. However, some of such techniques require raw US frequency data which is not accessible for all commercial machines. The accuracy obtained from such techniques is  poor compared to the tracked probe approach \cite{mercier2005review}. The accuracy of such phantoms rely on how well the phantom feature can be localized in B-scan as well as in the phantom.  
 
 \section{Stylus based techniques}
Stylus techniques \cite{khamene2005novelphatomless, hossbach2013simplified,hsu2009freehand} also referred to as Phantomless techniques \cite{khamene2005novelphatomless} use a stylus pointer or a needle \cite{hossbach2013simplified} instead of a calibration phantom. A stylus uses a localizer to calibrate itself by rotating its tip then used for ultrasound calibration. Moreover, stylus calibration contributes to error. In the technique by Hossbach et. al\cite{hossbach2013simplified}, stereo cameras attached to the probe and a needle are used to find the pose of the  US plane. Using a set of lines and corresponding intersection points, the calibration procedure is performed. 
%\section{Calibration Systems} %wording
\section{Camera based pose estimation}
Most widely used optical tracking systems such as Polaris and Optitrack with reflective markers provide sub-millimeter accuracy. However, such systems are very expensive and bulky. The electromagnetic sensor based systems are susceptible to the environment noise and cannot be used with metallic objects (surgical instruments) nearby. Moreover, all these techniques need special conditions to ensure the pose estimation. 
To overcome these problems, the researchers have been actively developing cost effective and reliable techniques. 

Many researchers \cite{hossbach2013simplified,mbaba2016low} have proposed low cost tracking systems with stereo-camera on the probe itself. Although these systems\cite{mbaba2016low,sun20136} avoid the problem of line of sight in other optical techniques, they use artificial skin markers to locate the cameras. Such markers can be problematic during clinical procedures, especially if they must remain from one scan to the next. In \cite{sun20136}, the authors proposed Simultaneous Localization and Mapping (SLAM). The skin markers are used to estimate the pose with respect to the cameras mounted on the probe.

\section{Least-squares solution}
Several methods have been proposed over the years to solve the probe calibration problem. As there is no definitive solution to the calibration problem, the minimum residual error (in a least-square sense) is used to solve the overdetermined system. Most methods either use iterative or non-iterative (closed form)\cite{mercier2005review}. For the iterative methods, the phantom features are generally unknown \cite{mercier2005review}. Iterative Closest Point (ICP) and Levenberg–Marquardt algorithms are frequently used in the literature. Most calibration problems are resolved by either iterative or closed-form solution. However, the solution depends on the phantom geometry and overall setup of the system. The wall phantoms, for example, cannot be solved with closed-form solutions because it is not possible to find the exact solution of the line defined by the wall\cite{mercier2005review}.

\section{Summary}
The probe calibration problem has been studied over the years for different objectives like 3D volume reconstruction, image guided therapies or image registration. All the techniques proposed are based broadly similar to the methods described in this chapter. The different setups and phantoms are determined on the objectives of the problem being solved. In our approach, we are using the solution of the probe calibration to register US images with MRI/CT. The proposed method is explained in the following chapter.
%\subsection{Systems using Optical markers}
%\subsection{Systems using Image Registration}
%\subsection{Systems using Stylus}
%\subsection{(Later on I will update this subsection)}
%\subsection{Temporal calibration ?}
%\subsection{Systems using Electromagnetic sensors}
%\subsection{Iterative Solution}
%\subsection{Closed-form solutions}

\chapter{Methodology} \label{chap:Methodology}

\section{Introduction} \label{sect:methodintro}
In this chapter, we describe the details of the proposed calibration method. While several calibration methods have been proposed over the years as discussed in Chapter~\ref{chap:StateofArt}, the general procedure for the calibration is similar \cite{mercier2005review}. The procedure can be divided into three steps: 1) pose estimation of the probe, 2) US image acquisition and feature extraction, 3) computation of the transformation respectively. We use optical tracking to estimate the pose of the probe. The US feature extraction is then performed which depends on the phantom landmarks used. The required transformation is subsequently calculated using least-square minimization.

The calibration task is achieved by rigidly attaching a pose tracking sensor to the US probe and scanning the calibration phantom. A tracking device provides the position and orientation in the sensor co-ordinate system. In our approach, optical tracking with different markers was used to estimate the pose of the phantom and the probe. As discussed in Chapter \ref{chap:StateofArt}, most of the techniques differ in terms of the calibration phantoms used. In this work, three different phantoms were designed and tested for the calibration procedure.   
\section{Theoretical consideration} \label{sec:theory_method}
\par
\begin{figure}[htbp]
\centering
\includegraphics[scale=0.5]{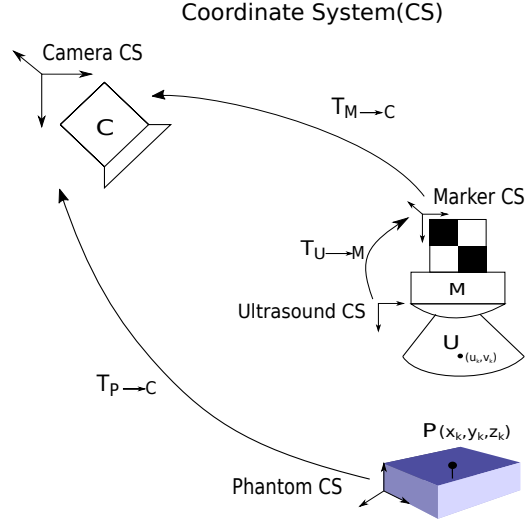}
\caption{Graphical representation of co-ordinate systems}
\label{fig:calib_repre}
\end{figure}

 The freehand probe calibration (also referred as ultrasound calibration) typically consists of several co-ordinate system (CS)'s to estimate the US imaging plane in the space. The total system and the relationships of the different CSs are shown in Figure~\ref{fig:calib_repre}. 
The aim of the probe calibration is to estimate the spatial relationship between US image and the tracking sensor $M$ attached to the probe. Four CS's are represented in the system. The optical tracking device (camera) was considered as the reference CS (also referred as world CS). All the transformations are estimated from the world CS or in other words from camera CS. In every frame captured, the pose of the probe represented with $T_{M\rightarrow C}$ was ascertained. Moreover, $T_{P\rightarrow C}$ was also measured with the camera at world CS. 
The calibration problem can be formulated with the Eq~\ref{eq:main}. 
The problem involves mapping the B-scan co-ordinates $(u, v)$ of the phantom features into the Marker CS. The aim of the calibration procedure is to estimate the rigid transformation ($T_{U \rightarrow M}$) between the US image CS ($U$) and the tracking device CS ($M$) provided that the phantom does not move during calibration. 
 
\begin{equation}
\begin{pmatrix}
x   \\
y \\ 
z \\
1
\end{pmatrix}
 = T_{P\rightarrow C} T_{M\rightarrow C}  T_{U\rightarrow M}  \begin{pmatrix}
s_x  u_k \\
s_y  v_k \\
0 \\
1
\end{pmatrix}  %\]
\tag{2}
\label{eq:main}
\end{equation}

where $s_x, s_y$ are scaling factors to map pixels to millimeters, $(x,y,z)$ coordinates of points in the phantom CS.

 From the Eq~\ref{eq:main} the calibration problem can be restated as the set of feature points in US image with the coordinates $(u,v)$ and the corresponding feature points in the physical phantom space. Thus, a least-square minimization can be used to estimate the required transformation $T_{U\rightarrow M}$. This overdetermined system of equations can be solved using minimum residual error as there is no exact solution \cite{mercier2005review}. 

\section{Closed-form solution}\label{sec:closedformsol}
The required transformation between two CS was estimated using closed-formed solution given by Horn in \cite{horn1987closed}. It consists of a set of two point sets ${Q_k}$ of the phantom features and the set of corresponding US image feature points $P_k$. The phantom feature points in Phantom CS ($x_p, y_p, z_p$)' are mapped to the Marker CS ($x_m, y_m, z_m$)' by following equation: 

\begin{equation}
\begin{pmatrix}
x_m   \\
y_m \\ 
z_m \\
1
\end{pmatrix}
 = (T_{M\rightarrow C})^{-1} T_{P\rightarrow C}  \begin{pmatrix}
x_p \\
y_p \\
z_p \\
1
\end{pmatrix}  %\]
\tag{3}
\label{eq:phantom2US}
\end{equation}

 Horn presented the solution to this least-squares problem of absolute orientations using unit quaternions in \cite{horn1987closed}:
\begin{equation}
\underset{R,t,s}{min} \sum^N_{k=1} \parallel Q_k - s\hat{\textbf{R}} P_k  - \textbf{t} \parallel^2_2
\tag{4}
\label{eq:horns_eq}
\end{equation}

where $\hat{\textbf{R}}$ is a rotation represented by unit quaternions, $\textbf{t}$ is the 3D translation vector which maps ${P_k}$ to ${Q_k}$, $\textbf{s}$ is the scaling factor,  $P_k$ is set of points in Marker CS obtained from Eq~\ref{eq:phantom2US}. $N$ is the number of images captured from different viewpoints. The required transformation is rigid-body transformation with  6 degrees of freedom (DoF). It involves 3-DoF for translation, 3-DoF for rotation (direction of axis of rotation and angle of rotation). Three points in both co-ordinate system provide nine constraints. Thus, three points in each CS are enough to find the transformation. We can determine the scaling factor from the US image which converts pixels to $mm$. If we do not consider the scaling factor, the transformation becomes the similarity transformation. This adds to one another DoF and total DoF becomes 7. The Horn's method also provide the solution for the similarity transformation. 

 The algorithm works in three steps. 1) It aligns all the point coordinates relative to the centroids. 2) Rotation is then computed by calculating eigen-vector corresponding to maximum eigen-value. 3) Once the rotation is known, the translation is calculated by rotating the centroid of ${P_k}$ with $\hat{\textbf{R}}$ given in Eq~\ref{eq:translation}: 
 \begin{equation}
\textbf{t}= Q_k - \hat{\textbf{R}} P_k
\tag{5}
\label{eq:translation}
 \end{equation}

\section{Experimental Procedure}\label{sec:exp_proc_steps}
\begin{figure}[!htb]
\centering
 %\framebox(200,200){Probe Casing}
\includegraphics[width=7.5cm,height=5.5cm]{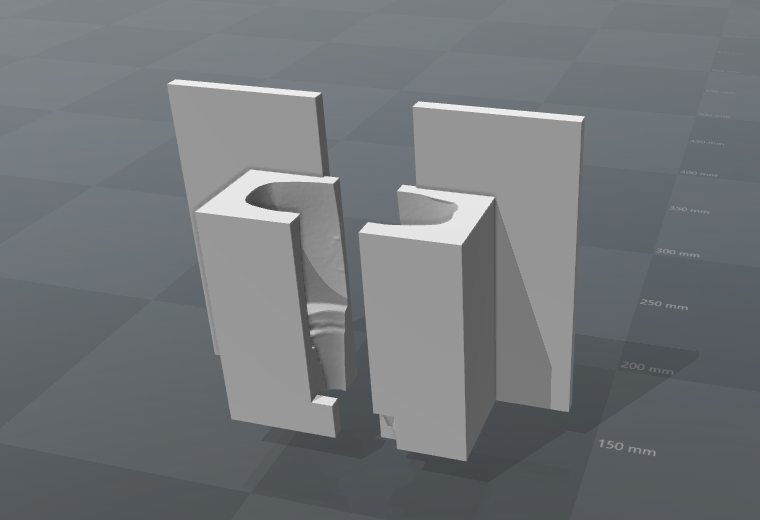}
\caption{3D printed probe casing} 
\label{fig:Probe_CasingCAD}
\end{figure}

The calibration system is realized with a monocular camera and a calibration phantom with known geometry. The probe casing (Figure~\ref{fig:Probe_CasingCAD}) was designed to rigidly attach the marker to the probe. It was designed by 3D scanning of the probe and constructed using a 3D printer. A structured light 3D scanner (HP David Scanner, United States) was used to scan the probe. All Computer Aided Design (CAD) modeling was done in Autodesk 3DS Max. The models of the probe casing and the phantoms were 3D printed using Ultimaker Extended 2 printer. 

The overall setup of the proposed system is shown in Figure~\ref{fig:exp_setup}. The procedure of data acquisition was performed according to the following steps for all custom designed phantoms:

\begin{figure}[!htb]
\centering
\includegraphics[width=11cm,height=6cm]{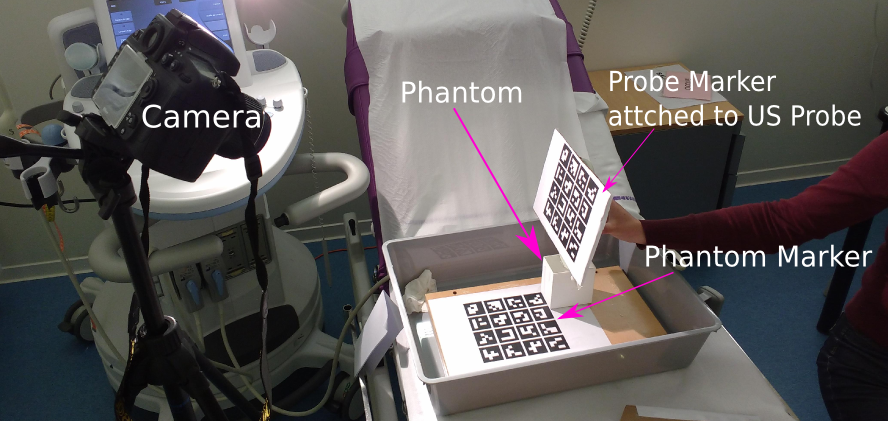}
\caption{Experimental setup} 
\label{fig:exp_setup}
\end{figure}

\begin{itemize}
\item The camera calibration was performed before the acquisition. It will be discussed in the Section~\ref{sec:cameracalib}.
\item The calibrated camera was positioned on a tripod in a manner that both the phantom and the probe lie in the field of view.
\item  The phantom marker was placed firmly in the container as shown in Figure~\ref{fig:exp_setup}. 
\item The origin of the Phantom CS (Figure~\ref{fig:calib_repre}) was considered at one of the corners of the phantom marker.
The phantom was then rigidly positioned relative to the origin of the Phantom CS.  The phantom features are then measured from the origin of the Phantom CS in the physical space (in world units $mm$). In other words, $x_p, y_p, z_p$ from Eq.\ref{eq:phantom2US} were measured in this step.  
\item The phantom container was filled with water to ensure coupling of the US beam.
\item The 3D printed probe casing with the probe marker was attached rigidly to the US probe (Figure~\ref{fig:exp_setup}).
\item Measurements were taken subsequently as explained in the following Subsection~\ref{sec:dataaqui}.
\item After the data acquisition step, a phantom feature detection algorithm was implemented to localize phantom features in B-scan. The details of the algorithm will be explained in the following Section \ref{sec:PhantomFeatureEx}.
\item Once all the correspondences are known, closed form solution described in the previous section was implemented to estimate the required transformation.

\end{itemize}

\subsection{US data acquisition}\label{sec:dataaqui}
The measurements were performed using the Aixplorer US device (SuperSonic Imagine, Aix-en-Provence, France) and the curved transducer XC6-1 (SuperSonic Imagine, Aix-en-Provence, France).  
\begin{itemize}
\item B-scan images of the phantom were acquired while sweeping the probe at different angles.
\item US probe was held steadily with least possible hand trembling. It was positioned in a manner that phantom features are clearly visible in the B-scan. 
\item Optical images and US images are then captured simultaneously by sweeping the probe. It was swept across the phantom with different angles and rotations. The accuracy is directly proportional to the number of measurements. The temporal calibration in US images is the synchronization between B-scan and the corresponding pose of the probe. In the proposed system, the temporal calibration is achieved by capturing the B-scan and clicking the camera shutter simultaneously.
\item Both B-scans and camera images were captured simultaneously to find the correspondences between them. In this way, temporal calibration is performed while acquisition.
\end{itemize}

\section{Synthetic data generation} \label{sec:syndata_method}
We estimate the system model by parameterizing the transformations with respect to the world CS (Figure~\ref{fig:calib_repre}). We parametrize the system shown in Figure~\ref{fig:calib_repre}. A single point phantom feature is considered to realize this system. The constraint to this problem is that the single point should be visible in the XY plane of the Marker CS. 

A fixed point $P(x,y,z)$ is considered in Camera CS. Camera CS is fixed at the origin in the space. All the transformations are measured with respect to the camera. The problem can be formulated such that the imaging plane defined by marker CS intersects point P.  Let us consider $T_{P\rightarrow C}$ and $T_{U\rightarrow M}$ are known and set to the appropriate values. Moreover, $P_k$ is randomly generated. Now, the problem can be reformulated using standard transformation equation \[X^{'}=Rx+t\] 
We can parametrize translation $t$ of the marker CS in a predefined range of $R$ as defined in Eq.~\ref{eq:syn_data_eq}.

\begin{equation}
t = T_{P\rightarrow C}\begin{pmatrix}
x_p   \\
y_p \\ 
z_p \\
1
\end{pmatrix}
 - R(R_{U\rightarrow M} P_{k} +t_{U\rightarrow M});
 \tag{6}
\label{eq:syn_data_eq}
\end{equation}
The generated synthetic data is shown in result Section~\ref{sec:syntheticdataresults}.

\section{Phantom design}
\par In Chapter \ref{chap:StateofArt} we discussed different types of phantoms, which provide variable accuracy, calibration speed and use different data acquisition techniques \cite{hsu2009freehand}. Although all the phantoms are geometrically different, they have a common property that they are constructed and placed in a container filled with a coupling medium, mostly water \cite{mercier2005review}. The phantom should be designed according to the objectives of the system. In this work, our objective was to keep the phantom design simple and to support automatic feature detection to ease the process. The general calibration for all the phantoms is similar. In this chapter, we discuss the phantom design considered for the project. While three different phantoms were designed, only the final design was used for the calibration.
\subsection{4-wire Phantom}
\begin{figure}[!htb]
\centering
\subfloat[Top view]{
\includegraphics[width=6cm, height=4cm]{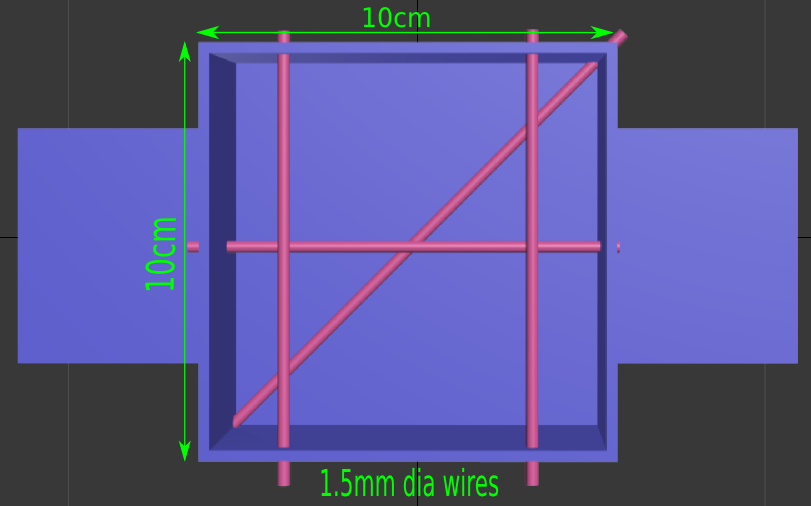}
}
 \subfloat[Side View]{
\includegraphics[width=6cm, height=4cm]{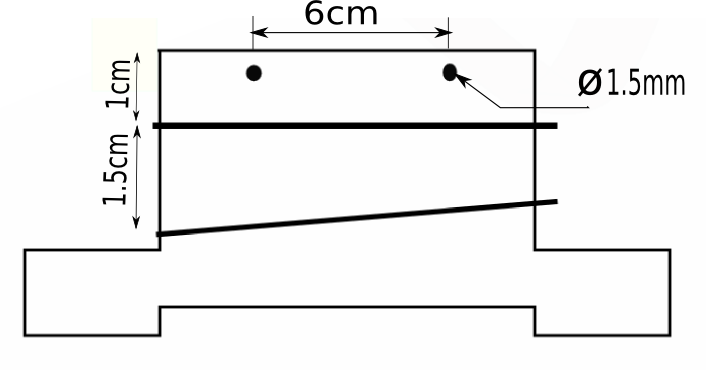}
}

\subfloat[3D printed phantom]{
\includegraphics[width=6cm, height=4cm]{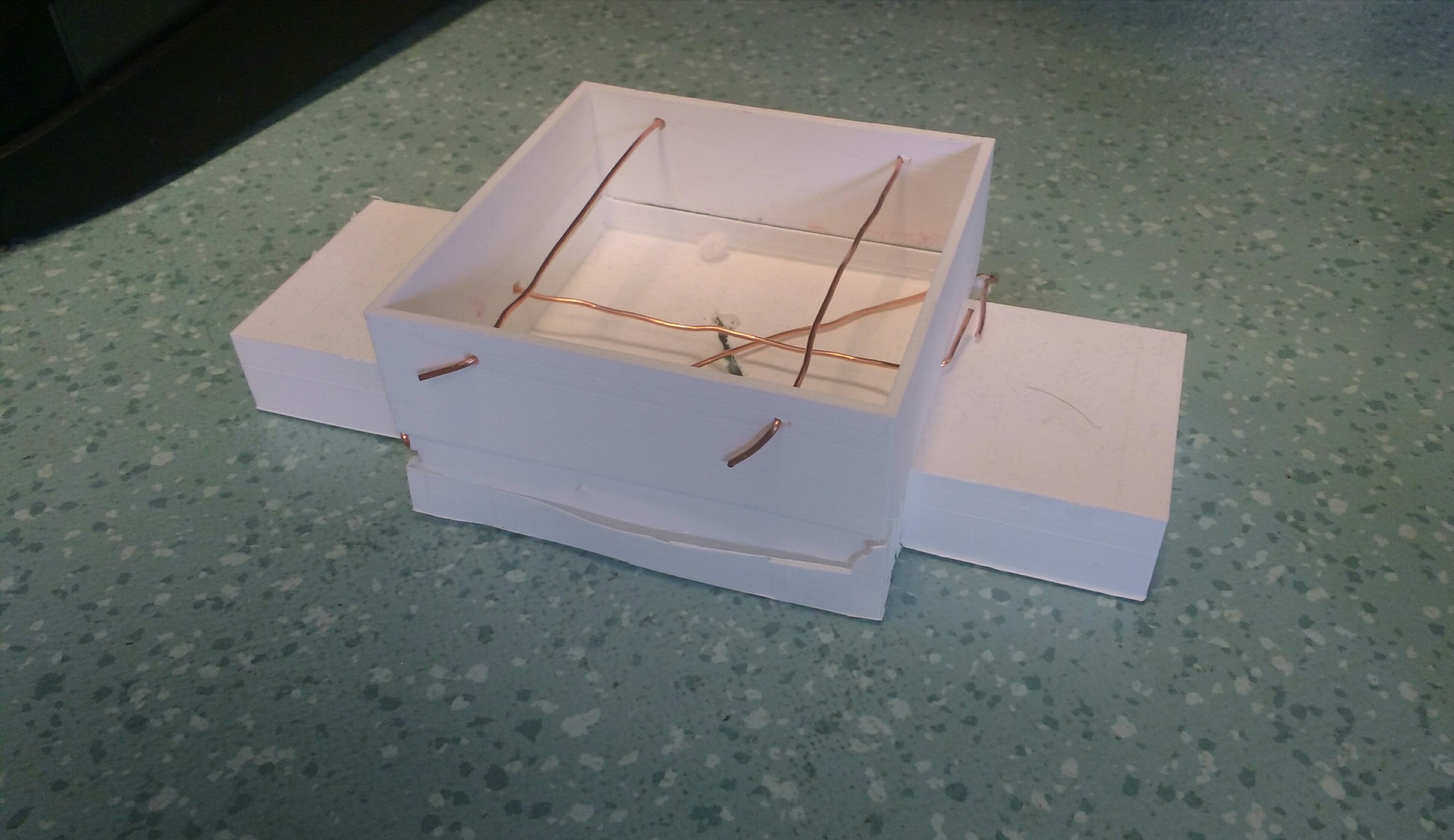}
}
\caption{4-wire phantom} 
\label{fig:4wireph}
\end{figure}

The first phantom was composed of two cross wires and two parallel wires as shown in Figure~\ref{fig:4wireph}. This phantom was designed to incorporate the automatic segmentation. The wires in this phantom are arranged to provide more features in all the axes with less number of scans required. The phantom  was built using 3D printing and four copper wires with diameter $1.5$ $mm$. The geometry of this phantom composed of a box and two supporting structures (Figure~\ref{fig:4wireResult}). As the 3D printed material floated on water, the support structures were designed to stabilize it underwater using support-weights. The B-scans for this phantom will be discussed in Section~\ref{sec:result4wire}. This phantom was rejected due to the following reasons. First, multiple echoes were received in the B-scan. Moreover, the dimensions of the phantom were large, so it was difficult to see all the wires simultaneously in a single B-scan. Second, the geometry of the phantom did not allow to hold the probe steadily as there was no resistance in the water. 
%\begin{figure}[!htb]
%\centering
%\includegraphics[scale=0.08]{setup.jpg}
%\caption{Experimental Setup}
%\end{figure}
\subsection{Point Phantom}\label{sec:PointPhDesign}
 Point phantoms are simple and easy to construct comprising a single point which can be imaged from different poses. For accurate calibration, sufficient measurements have  to be acquired from diverse viewpoints \cite{mercier2005review}. The US probe has to be aligned in order to scan the single point accurately. Calibration can be time-consuming with such phantoms. The automatic detection of single points in the B-scan is not reliable. Therefore, the calibration process require manual segmentation which makes it more tedious.
\par
\begin{figure}[!htb]
\centering
\includegraphics[width=6cm, height=4cm]{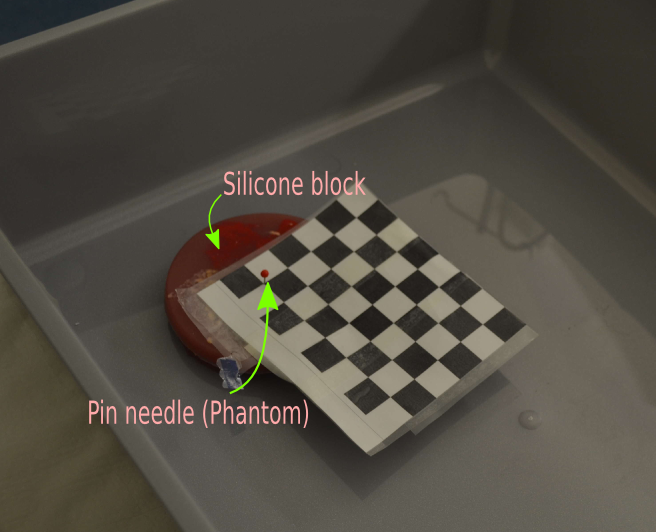}
\caption{Point phantom}
\label{fig:point_ph}
\end{figure}

 A plastic pin head needle was used as a point phantom. Figure~\ref{fig:point_ph} shows the pin needle attached to the silicone block. It was then immersed in a water bath (coupling medium). The B-scans with this phantom were unreliable consisting of artifacts. The results will be discussed in Section~\ref{sec:point_result}. 

\subsection{Hemisphere Phantom}\label{sec:hemi_design}
\begin{figure}[htbp]
\centering
\includegraphics[width=9cm, height=6cm]{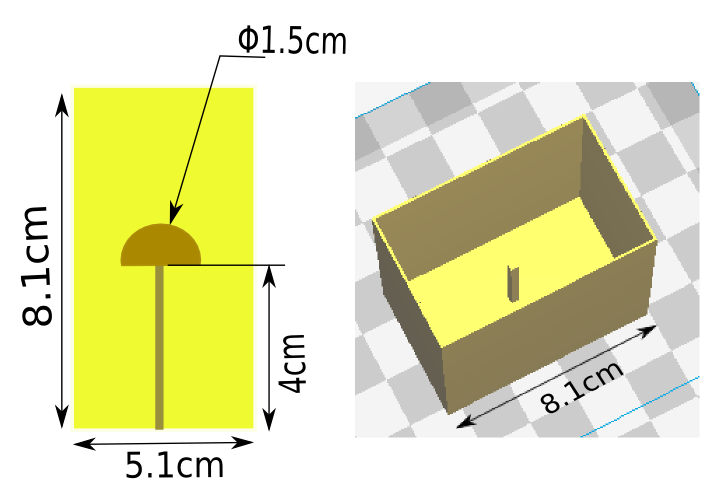}
\vspace{2mm}
\includegraphics[width=6cm, height=8cm]{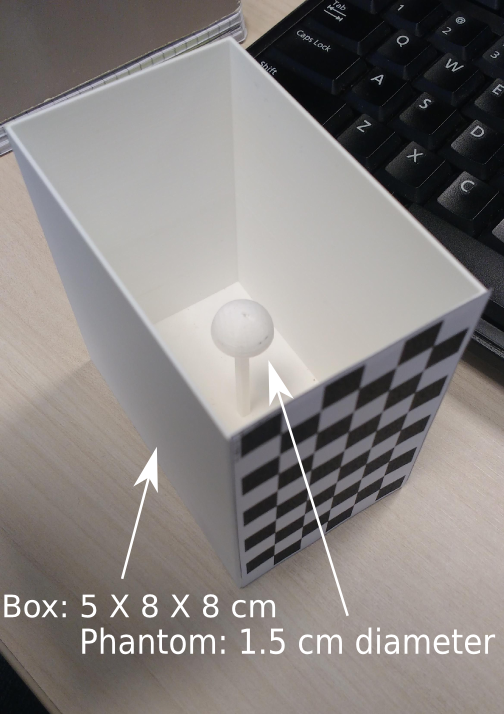}
\caption{Hemisphere phantom and its CAD Model } 
\label{fig:hemi_phantom}
\end{figure}

 A spherical object gives a contour (circle) in B-scan to allow robust and automatic segmentation of landmarks in B-scans \cite{bakulina2016automated}. This type of phantom can be categorized as a point phantom because only a single point (sphere's center) is considered as a landmark/feature. Other research groups\cite{bakulina2016automated, brendel2004simple} have used similar approaches of using spherical objects as a phantom due to their ease in fabrication. A similar approach was implemented in this work. An arc appears in the B-scan when any spherical object is scanned. Thus we can just use a hemisphere object because only half part of the sphere is visible in the B-scan. The hemisphere was positioned at the center of the bounding box. A safe distance from the bottom of the container has to be maintained to be able to exclude artifacts received from the bottom of the container. Therefore, we designed a hemisphere with a pillar which cannot be seen in the B-scan. 
 The geometry of this phantom was designed such that the transducer was supported by the walls which allows to hold it steadily. In the previous 4-wire phantom, we noticed that the 3D printing material was clearly visible in the B-scan. Hence, we 3D printed the hemisphere using the same material. The geometry of the phantom is shown in Figure~\ref{fig:hemi_phantom}

\chapter{Implementation} \label{chap:implementation}
This chapter describes the implementation details of the proposed pose estimation and landmark detection algorithm. We use a single calibrated camera to estimate the pose of the probe with 6-DoF. The US feature extraction algorithm is then performed. The required transformation is subsequently calculated using least-square minimization as described in Section~\ref{sec:closedformsol}.

\section{Camera Calibration} \label{sec:cameracalib} %Wordin Technical background
The camera calibration is performed to calculate camera parameters: intrinsics, extrinsic and distortion coefficients. The intrinsic parameters $K$ include focal lengths ($f_x,f_y$) and the optical center ($c_x,c_y$) of the camera. The extrinsic parameters include orientation $R$ and position $t$ of the camera. A pin hole camera model is used to map the 3D world points to 2D image points with projection matrix $P$ \cite{bouguet2004camera}. The algorithm uses the size of the checkerboard pattern in world units to find the correspondences between 3D world and 2D image plane. The projection from 3D to 2D is given by P.
$$ 
K=\begin{bmatrix}
 f_x & 0 & c_x \\
& f_y & c_y \\
&  & 1 \\
\end{bmatrix}
$$

\[P=K|R t\]

\begin{figure}[!htb]
\centering
\includegraphics[width=6cm,height=6cm]{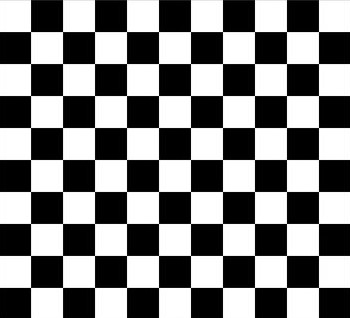}
\caption{Checkerboard pattern} 
\label{fig:Checkerboard_pattern}
\end{figure}

 The planar fiducials (Figure~\ref{fig:Checkerboard_pattern}) are used as correspondences. The camera parameters are estimated by taking multiple images of the checkerboard pattern (Figure~\ref{fig:Checkerboard_pattern}) from different view points and orientations. For each view, individual extrinsic are present while intrinsics remain the same for all views.  In our checkerboard approach described in Section~\ref{sec:checkerboard_ex}, the extrinsic (camera pose) from the camera calibration is used to estimate the pose of the probe marker. The extrinsic provide 6-DoF with rotation and translation.

 There are many camera calibration algorithms available. However, we used the open source MATLAB toolbox by Bouguet et al\cite{bouguet2004camera}.
 
%\section{Temporal Calibration(Optional section)}
%\textcolor{red}{talk about temporal calibration which is to %synchronize images taken and KLT approach tried}
%{KLT Tracking Approach:Optional}

\section{Pose estimation}
\begin{figure}[!htb]
\centering
\includegraphics[width=8cm, height=6cm]{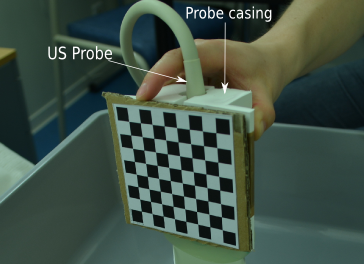}
\caption{US probe with 3D printed casing to hold marker}
\label{fig:USprobe}
\end{figure}

Obtaining the
camera pose from images requires to find the correspondences
between known points in the environment and their
camera projections
The transformations needed for the calibration were calculated by the pose estimation of the probe. The pose estimation requires the correspondences between the world points and the projections. While there are several methods to solve this problem, the fiducial markers are the most commonly used technique because of the high precision and high accuracy gained \cite{munoz2012aruco}. In our approach, the single camera system with fiducial markers were used to estimate pose of the probe. The first setup involved estimating the pose from a planar checkerboard. Moreover, we exploited the performance of pose estimation using Augmented Reality (AR) markers.
The fiducial markers are rigidly attached to the probe using a 3D printed  probe casing (Figure~\ref{fig:USprobe}, Figure~\ref{fig:Probe_CasingCAD}).
The attached marker represents Marker CS. The aim of the calibration is to estimate the transformation between this marker and the US imaging plane.

%\subsection{Introduction}
\subsection{Checkerboard Marker System}\label{sec:checkerboard_ex}
A marker rig was used to estimate both poses of the phantom and the US probe. Two checkerboards with similar dimensions, one for the phantom  and another one for the probe were used. One checkerboard is attached to the container in which the phantom was placed and another checkerboard was attached to the probe casing. The camera calibration algorithm described in Section~\ref{sec:cameracalib} provides the extrinsics of the camera or in other words the transformation of both the checkerboards with respect to the camera CS.
We obtain both  $T_{P\rightarrow C}$ and  $T_{M\rightarrow C}$ using the camera calibration algorithm. The important thing to consider here is that the origin of the Marker CS should be fixed at one of the corners of the checkerboard for all poses of the probe.

\subsection{AR Marker System}
\begin{figure}[htbp]
\centering
\includegraphics[width=7cm, height=7cm]{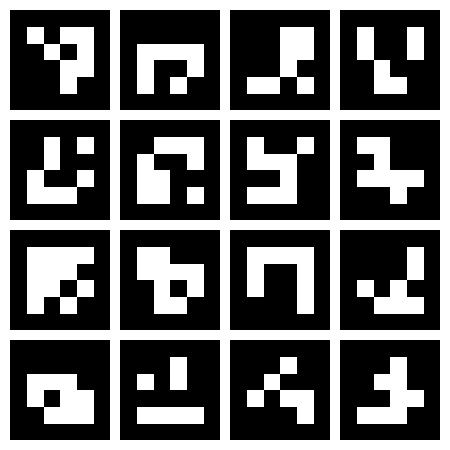}
\caption{ArUco example board of markers}
\label{fig:ArucoExample}
\end{figure}

Another set of marker rig was exploited for the pose estimation. 
These set of AR markers provide robust capabilities of real-time operation. ArUco markers\cite{munoz2012aruco} were used in this approach. ArUco library built with OpenCV provides high precision pose estimation \cite{munoz2012aruco} because the markers are easy to detect in the environment. It provides robust marker detection and error correction capabilities in different lighting conditions \cite{munoz2012aruco}. Each marker (Figure~\ref{fig:ArucoExample}) provides 3D pose of the camera with respect to each marker.

 Each marker is encoded with unique identification using predefined ArUco dictionary. Instead of using single marker, the plane of ArUco markers was used for better accuracy. The plane of markers also provide robustness to avoid occlusion issues with the single markers. 
 
  A $4 \times 4$ plane of markers was generated using a default dictionary (Figure~\ref{fig:ArucoExample}). ArUco estimates the camera pose from the four corners of the marker, given that the camera is calibrated. This approach is similar to the checkerboard approach described before. One plane of markers is attached to the probe (Probe Marker) and another one is used as the Phantom Marker. The overall system is shown in Figure~\ref{fig:exp_setup}.
  
 The ArUco system works in the following manner.
 The process can be divided in three parts: 1) Marker detection 2) Marker Identification and error correction 3) Pose estimation.
\begin{itemize}
\item Marker detection: Firstly, the image segmentation algorithm is performed on the gray-scale input image. A local adaptive thresholding approach is used for the extraction of contours. This approach is effective in different lighting conditions. 
Subsequently, the system detects all unique markers in the image using contour detection. The polynomial approximation is performed to obtain only the required markers from the image. The markers that are not approximated to 4-vertex polygons are discarded. The detected polygons from this process are shown in Figure~\ref{fig:aruco_algo}. 
\begin{figure}[htbp]
\centering
\includegraphics[width=10cm, height=5cm]{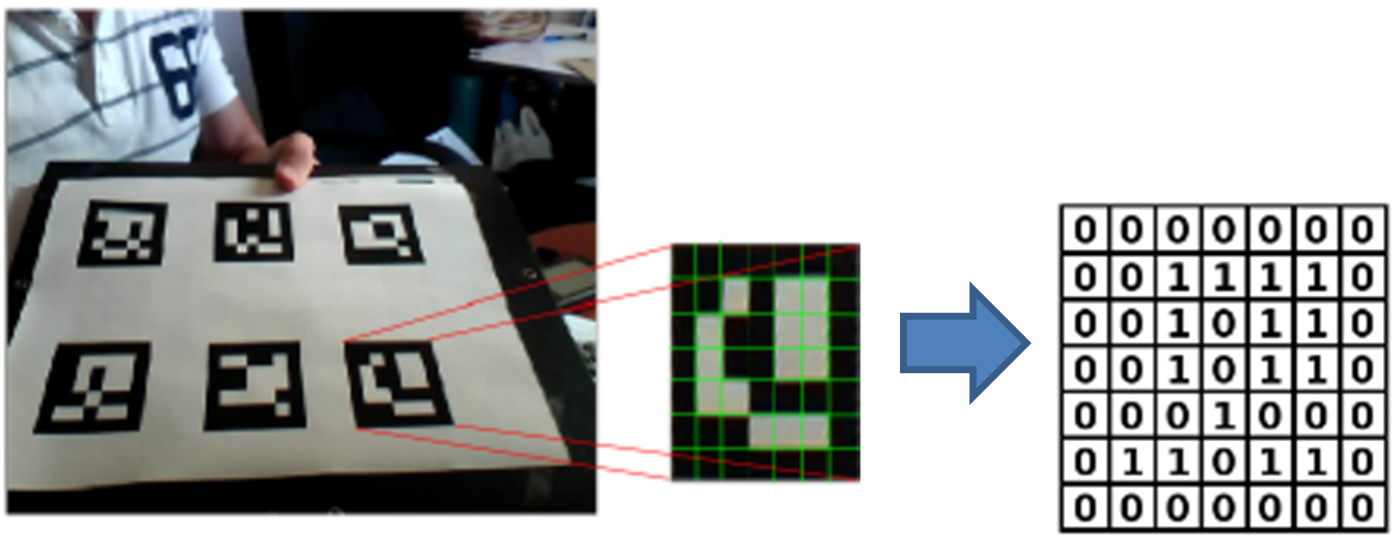}
\caption{Detected ArUco Marker after perspective transformation \cite{munoz2012aruco}.}
\label{fig:aruco_algo}
\end{figure}
\item Marker code extraction: Once all the required contours are extracted, they are analyzed to extract the internal code of them.
The homography matrix is computed to remove the perspective projection (Figure~\ref{fig:aruco_algo}). A optimal thresholding (Otsu's method) is performed to get binary image. The resulting image is divided into grid and assigned 0's and 1's to each element (Figure~\ref{fig:aruco_algo}).
The marker identification is then performed to determine which dictionary the detected marker belongs to. If more than one markers are used, the inter-marker distance is used to correct the erroneous markers candidates. 
\item Pose estimation: The corners of the detected marker/s are refined using linear regression of the markers side pixels to calculate the intersection. Using the refined corners, the pose is estimated by minimizing the reprojection error of the corners iteratively using Levenberg-Marquardt \cite{munoz2012aruco}. 

\end{itemize}

\section{Phantom feature extraction}\label{sec:PhantomFeatureEx}
\begin{figure}[!htb]
	\centering
	\includegraphics[width=8cm, height=6cm]{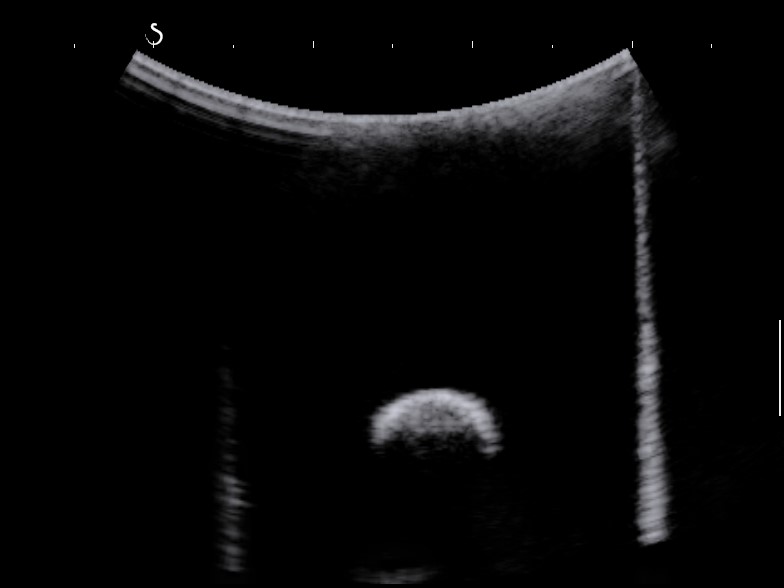}
	\caption{B-scan of Hemisphere phantom}
	\label{fig:hemi_bscan}
\end{figure}

The B-scan of a hemisphere phantom is an arc as shown in Figure~\ref{fig:hemi_bscan}. The required feature is the centre of the arc. It can be detected by using Hough Transform. 

 In Cartesian coordinate, the circle is describe by the following equation: \[(x-a)^2+(x-b)^2=r^2\] where $(a,b)$ is coordinates of the center and $r$ is the radius. The Hough transform considers edge map of the image. It maps each edge pixel into all parameter points which lie on the surface of an inverted right angled cone whose apex is at $(x, y, 0)$ \cite{illingworth1987adaptive}. 
The circle parameters are estimated by the intersection of many conic surfaces that are defined by points on the 2D circle. The problem of circle finding is resolved with two steps. The first step involves a two parameter hough transform to find the circle center while the second step is a simple histogramming to identify the radius, \(\delta=(x-a_0)^2+(y-b_0)^2\) where $(a_0,b_0)$ are centre co-ordinates estimated from first step. The highest peak in the histogram $\delta$ is identified as the radius \cite{illingworth1987adaptive}.

\chapter{Results} \label{chap:ExpAndResults}
%\section{Introduction}
In this chapter we present the experimental results of the proposed system. They are categorized based on different phantoms, pose estimation techniques and setups performed in this work.  
Moreover, we present the results of the initial phantom designs and why they were rejected for the calibration task.
The calibration results are presented only for the hemisphere phantom design that was used. The experimental results were obtained by following the procedure explained in Section~\ref{sec:exp_proc_steps}. 
In the calibration procedure, each stage (phantom fabrication, image
acquisition, spatial co-registration, image processing,
formulation, and numerical optimization solution), contributes to the error \cite{boctor2003rapid}.
%\section{Experimental Setup}
\section{4-wire Phantom} \label{sec:result4wire}
\begin{figure}[!htb]
\centering
\subfloat[]{
\includegraphics[width=7cm,height=5cm]{figures/wire.jpg}}

\subfloat[]{
\includegraphics[width=7cm,height=5cm]{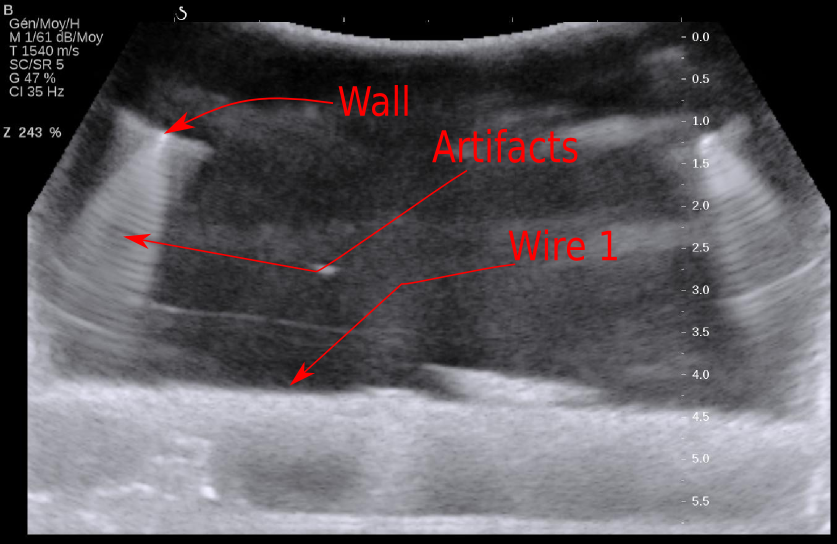}
}
\subfloat[]{
\includegraphics[width=7cm,height=5cm]{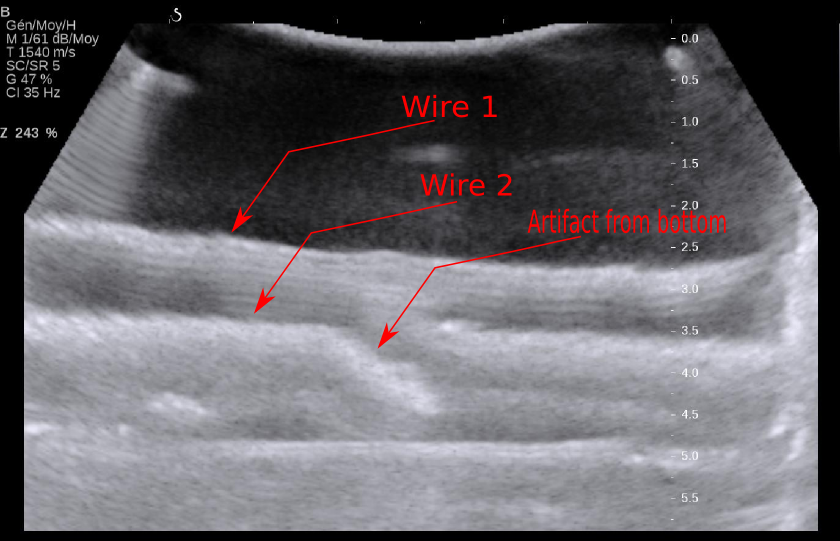}
}
\caption{a) 4-wire phantom, b \& c) Different B-scans of 4-wire}
\label{fig:4wireResult}
\end{figure}

 The first phantom was composed of wires. The experiment for this design was mainly performed to check if the phantom features are vividly available in the corresponding B-scans. The B-scans were obtained by immersing the phantom in water. The phantom and respective B-scans are shown in Figure~\ref{fig:4wireResult}. 
The B-scans (Figure~\ref{fig:4wireResult}) of this phantom included many artifacts. Moreover, not all wires and their intersections were vividly available in the B-scan. The presence of artifacts did not allow reliable implementation of an automatic segmentation algorithm. 

This phantom was rejected due to the following reasons. 1) Multiple artifacts were present in the B-scan which did not allow automatic segmentation. 2) The geometry of the phantom did not allow to hold the probe steadily as there was no resistance in the water. Moreover, it was not stable underwater. 3) The distance between parallel lines was greater than the size of the transducer. Thus, not all wires were not visible in a single B-scan. 4) If we assume the speed of the sound $c$ to be 1000 $m/s$ and US transducer to work in a frequency range $f$=1...6~$MHz$. The wavelength of US $\lambda = \frac{c}{f}$. This gives $\lambda \leq 0.2~mm$. Therefore, the wires with the thickness $\phi \geq \lambda$ can cause multiple echoes in the B-scan. Such echoes can be seen in Figure~\ref{fig:4wireResult} because we used the wires with thickness 1.5~$mm$. 5) Some of the artifacts occurred from the bottom of the 3D printed box due to the short height with respect to the transducer. Figure~\ref{fig:4wireResult}~c) also shows the artifact caused due the crack in the bottom of the 3D printed box.

\section{Point Phantom}\label{sec:point_result}
\begin{figure}[!htb]
\centering
\subfloat[]{
\includegraphics[width=7cm,height=5cm]{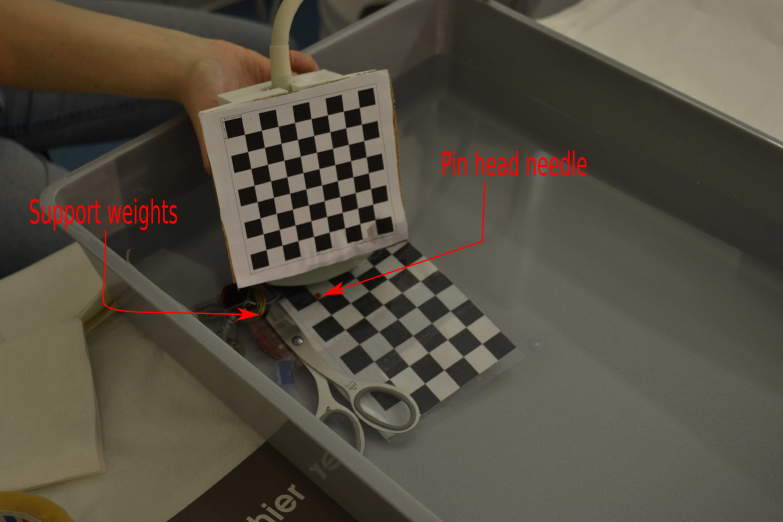}}
\subfloat[]{
\includegraphics[width=7cm,height=5cm]{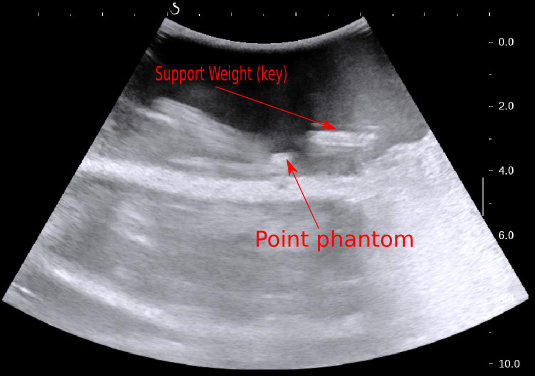}}

\subfloat[]{
\includegraphics[width=7cm,height=5cm]{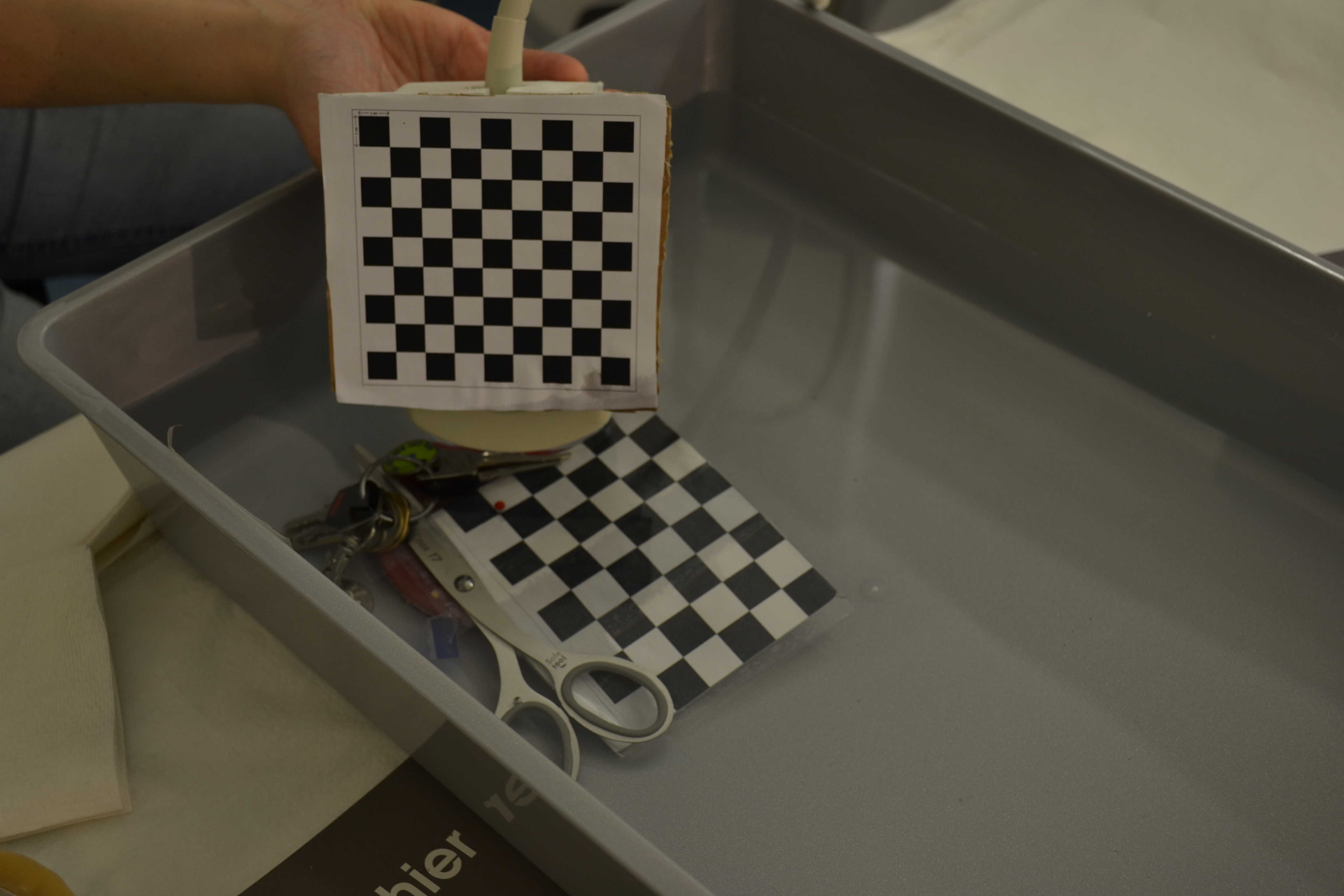}}
\subfloat[]{
\includegraphics[width=7cm,height=5cm]{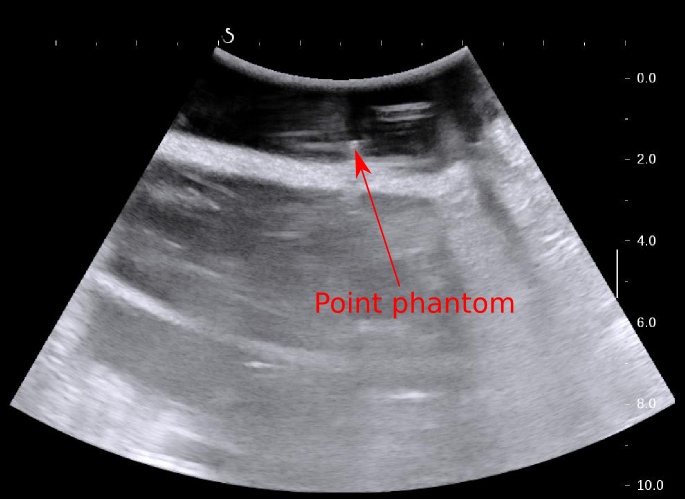}}
\caption{a) Point phantom pose 1, b) B-scan for pose 1, c) Point phantom pose 2, d) B-scan for pose 2}
\label{fig:PointResult}
\end{figure}

As described in Section~\ref{sec:PointPhDesign}, this phantom was design for its ease of construction. The pin head needle was inserted in the silicon block. The checkerboard was attached such that the needle was at one of the corners (origin of the board). The x-y coordinates of the point phantom were measured using the camera calibration. However, the z-coordinate of the phantom in phantom CS was measured manually as 12 $mm$. This assembly was immersed in the water to obtain B-scans as explained in Section~\ref{sec:dataaqui}. The resulted B-scans for different poses of the probe are shown in Figure~\ref{fig:PointResult}.

This phantom design was rejected due to following reasons. This phantom design produced many artifacts in the B-scan (Figure~\ref{fig:PointResult}). The automatic detection of the point is not reliable due to the artifacts present. Moreover, the silicone block used to position the needle was unstable underwater. Thus, we used some weights (keys, scissor Figure~\ref{fig:PointResult}(a)) to stabilize the phantom assembly. These weights resulted in some of the artifacts present in the B-scan. The bottom of the water container also produced some artifacts because the short height with respect to the transducer. 
\section{Hemisphere Phantom Results}\label{sec:hemi_result}
\begin{figure}[!htb]
\centering
\includegraphics[width=7cm,height=5cm]{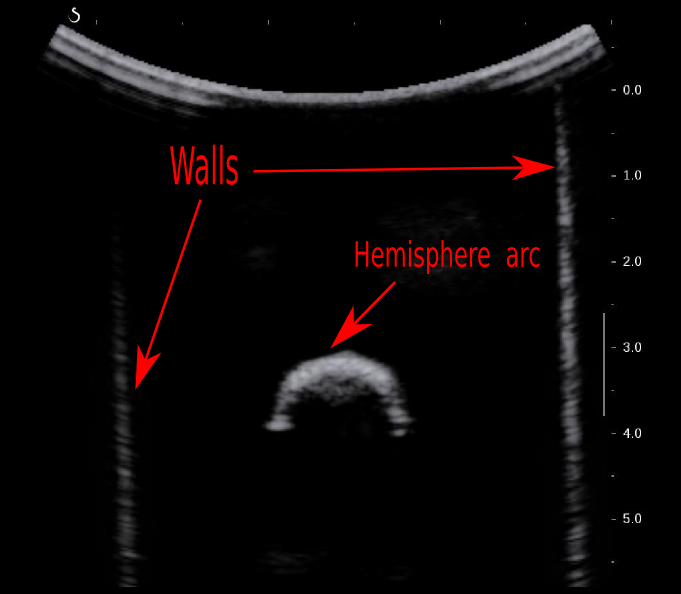}
\includegraphics[width=7cm,height=5cm]{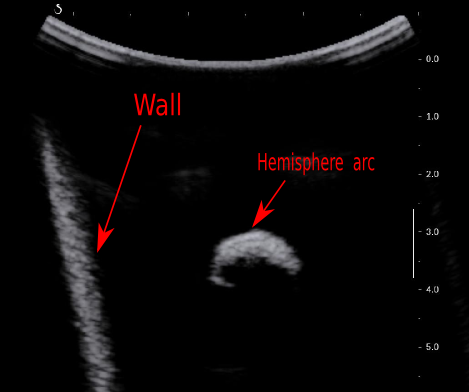}
\caption{Hemisphere phantom B-scans, left: probe vertically aligned, right: probe aligned with a tilt }
\label{fig:HemiResult}
\end{figure}

The problems faced with earlier designs of the phantoms were resolved with this phantom as discussed in Section~\ref{sec:hemi_design}. The B-scans (Figure~\ref{fig:HemiResult}) showed little or no artifacts present with this phantom. Some small artifacts were caused due to air bubbles produced while sweeping the transducer.

The hemisphere is clearly visible as an arc in the B-scan. Moreover, the walls of the phantom container are visible as straight lines. Although these lines can also be used as phantom features for the calibration, we only considered the center of the sphere as a landmark. 
This phantom was used for the calibration procedure. 

\subsection{Phantom landmark detection}\label{sec:phlandmarkdetectionresult}
\begin{figure}[!htb]
\centering
\includegraphics[width=7cm,height=5cm]{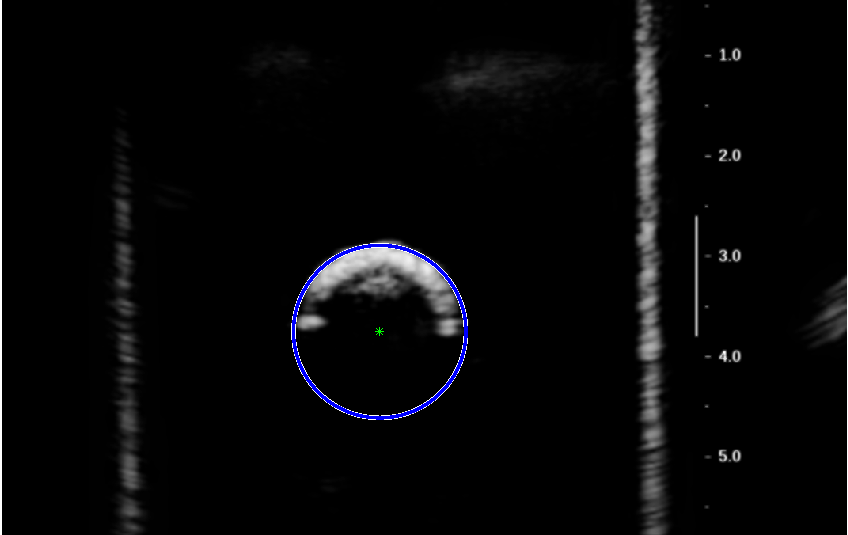}
\includegraphics[width=7cm,height=5cm]{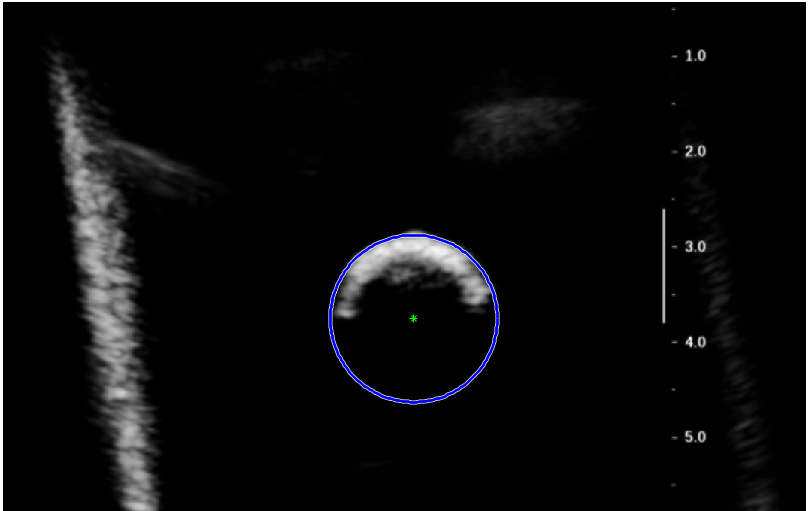}
\caption{Detected circle and its center of Hemisphere phantom }
\label{fig:Ph_Feat_Result}
\end{figure}

The hemisphere phantom used for the calibration task gives arcs in the B-scans. In our approach, we are interested in the center of this arc (center of hemisphere) as the phantom feature. This center detection was achieved using Hough Transform described in Section~\ref{sec:PhantomFeatureEx}. The detected center and the circle are shown in the Figure~\ref{fig:Ph_Feat_Result}. 
%Although we did not evaluate the accuracy of the center detection. We can evaluate the error in radius using the known scale factor (pixel to $mm$) of B-scan and real world distance of the hemisphere (15$mm$). We can see (Table~\ref{table:pha_land_detect}) that for some B-scans error is upto $1.5mm$ and minimum error is around $0.2mm$. Although with this error evaluation, we cannot assure that the center detection is wrong because circles can be concentric. However, we 

%\begin{table}[!htb]
%\centering
%\caption{Result table for error in radius}
%\label{table:pha_land_detect}
%\begin{tabular}{|c|c|}
%\hline
%Parameter & Erro in $mm$.\\ \hline
%Std.  & 0.256 \\ \hline
%Mean & 1.432 \\ \hline
%Min & 0.127 \\ \hline
%Max & 1.712 \\ \hline
%\end{tabular}
%\end{table}

\section{Synthetic data}\label{sec:syntheticdataresults}
\begin{figure}[!htb]
\centering
\includegraphics[width=11cm,height=10cm]{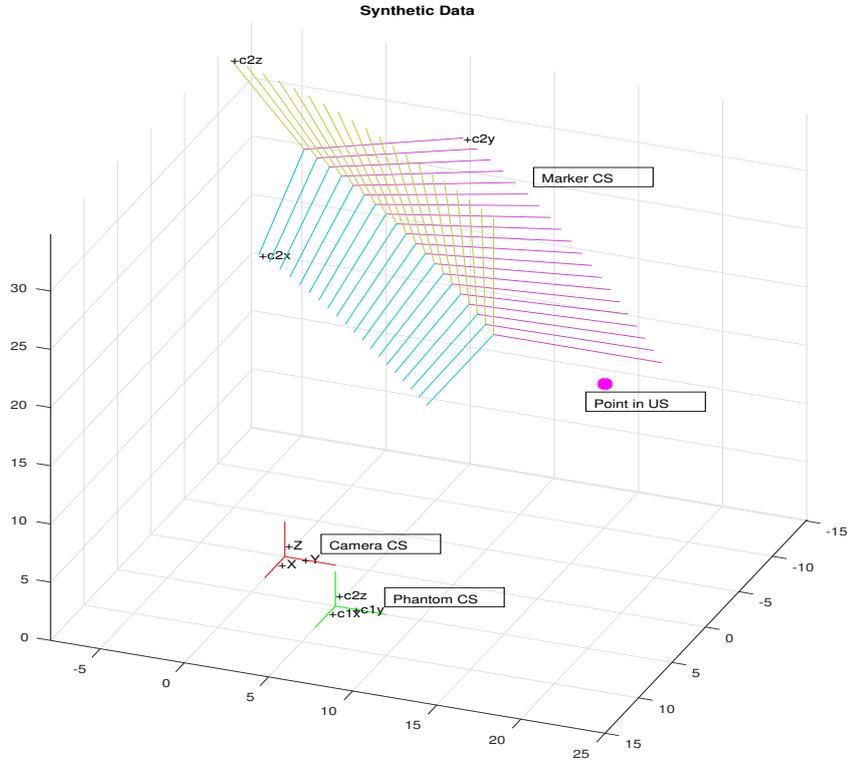}
\caption{The overview of all coordinate systems in the synthetic data}
\label{fig:SynResult}
\end{figure}

The synthetic data is generated to check the correctness of the closed-form solution. The 3D-plot of the overall CSs and the probe sweeping in the space with respect to the camera CS and phantom CS are shown in Figure~\ref{fig:SynResult}. As explained in Section~\ref{sec:theory_method}, the Marker CS was attached to the probe. Figure~\ref{fig:SynResult} shows the probe sweeping around $X$ and $Y$ axes with same rotation and fixed rotation along $Z$ axis using the synthetic data. In synthetic data generation (Section~\ref{sec:syndata_method}), the Camera CS, ultrasound CS and the Phantom CS are fixed. The related transformations to fixed CSs' were set to appropriate values to visualize. The aim is to determine the translation of Marker CS such that it intersects the Point in US (Figure~\ref{fig:SynResult}) using Eq.~\ref{eq:syn_data_eq}  as discussed in Section~\ref{sec:syndata_method}. 

In our approach, the effect of the Gaussian noise on the pose of the probe in the Marker CS is studied. Table~\ref{table:syn_data} shows this effect of varying $\sigma$ and zero mean. With no noise in the pose measurement, there was zero residual between two point sets. This error increased with increase in the noise. 

\begin{table}[!htb]
\centering
\caption{Error in synthetic data with varying noise }
\label{table:syn_data}
\begin{tabular}{|c|c|}
\hline
$ \sigma$ & Std.\\ \hline
0 & 0 \\ \hline
0.5 &   0.272 \\ \hline
1 & 0.504  \\ \hline
2 & 1.127   \\ \hline
\end{tabular}
\end{table}
   % 0.0000    0.0000    0.0000         0
   % 0.6157    0.6666    0.3913         0
   % 0.5534    0.6065    0.4049         0
   % 0.5098    0.5709    0.4375         0

%\section{Calibration results}
\section{Calibration results}
In this section we present the calibration results using the proposed approach with closed-form solution. The calibration procedure is performed only for the hemisphere phantom with two pose estimation techniques. The data acquisition for both approaches and all datasets was performed according to the steps discussed in Section~\ref{sec:exp_proc_steps}.
 The transformations $T_{P \rightarrow C}$ and $T_{M \rightarrow C}$ from Eq.~\ref{eq:main} are estimated using the camera extrinsics with respect to both the markers. The required transformation $T_{U \rightarrow M}$ is estimated using the closed-form solution. The calibration results were measured by transforming points from B-scan to the phantom CS using all the estimated transformations in Eq~\ref{eq:main}. We calculated and evaluated the results based on the Backprojection Residual Error (BRE).
The co-ordinates of the phantom feature (the center of hemisphere) $P(x,y,z)$ with respect to the origin of the phantom marker were known. The BRE is a measure of the absolute differences with the point P and re-projection of the corresponding point in the B-scan using Eq.~\ref{eq:main}. Ideally, the BRE should be zero for all set of images along all axes. \newline
\textbf{Note:} All the BRE values mentioned are measured in $mm$.

\subsection{Checkerboard system}\label{sec:checkerboard_result}
\begin{figure}[!htb]
	\centering
	\includegraphics[width=10cm,height=8cm]{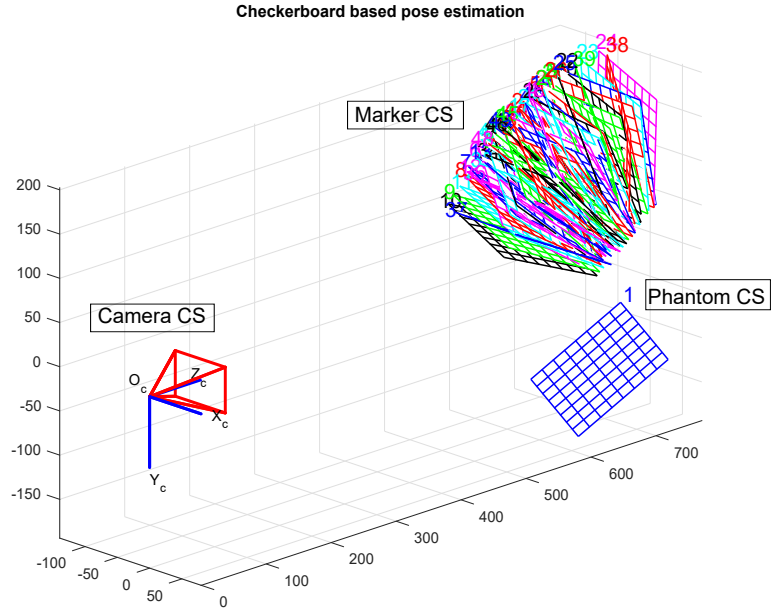}
	\caption{Different poses of the probe with respect to Camera and Phantom CS}
	\label{fig:checkpose}
\end{figure}

In this section, the results from the checkerboard system are described. Three measurements: Dataset 1, Dataset 2, Dataset 3 were acquired with this system of approach. All required poses were estimated using the Camera Calibration Toolbox by Bouguet et al.\cite{bouguet2004camera}. Figure~\ref{fig:checkpose} shows 3D visualization of different CSs and extrinsic parameters of camera. 

\subsubsection{Dataset 1}
\begin{figure}[!htb]
\centering
\includegraphics[width=7cm,height=5cm]{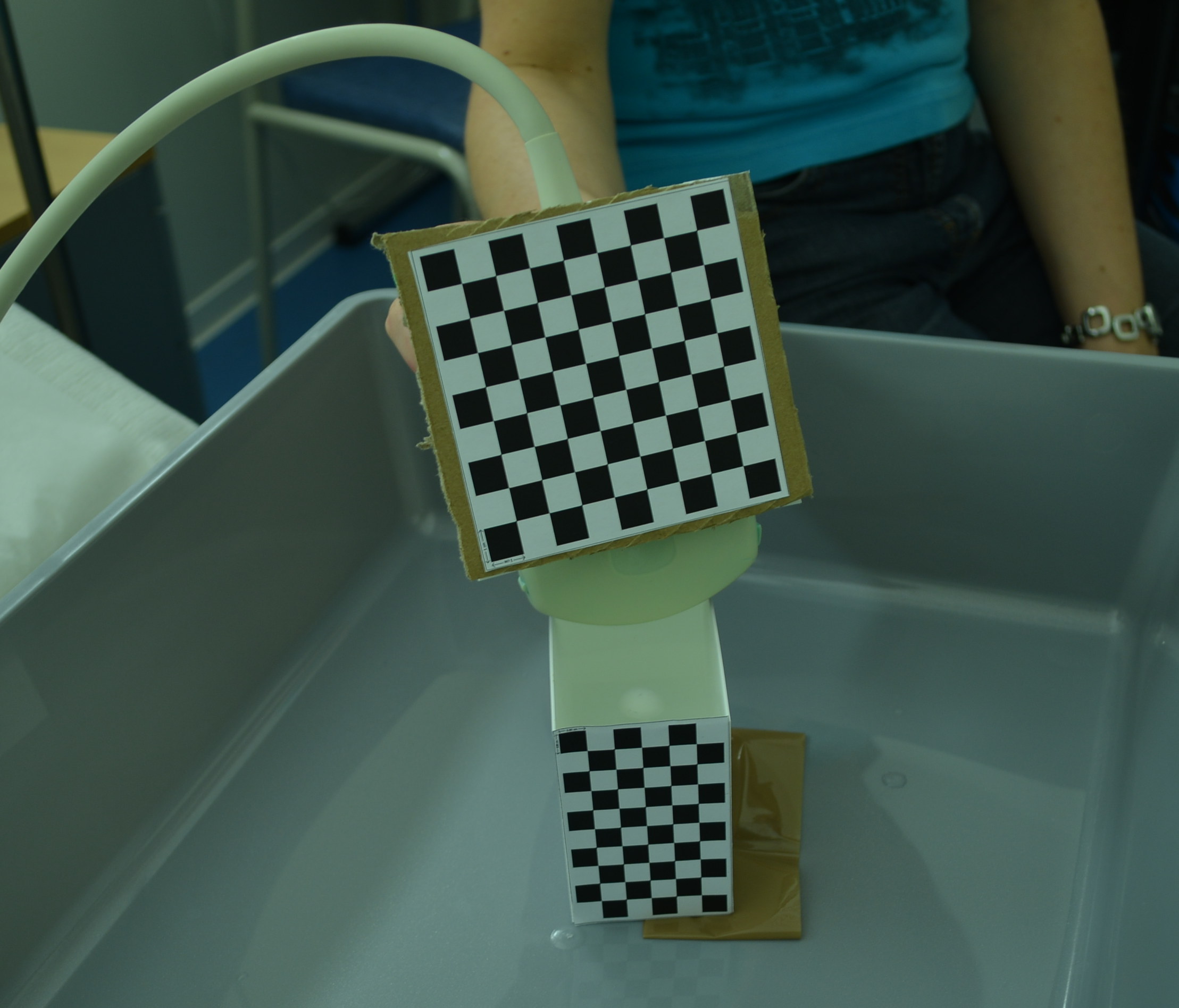}
\includegraphics[width=7cm,height=5cm]{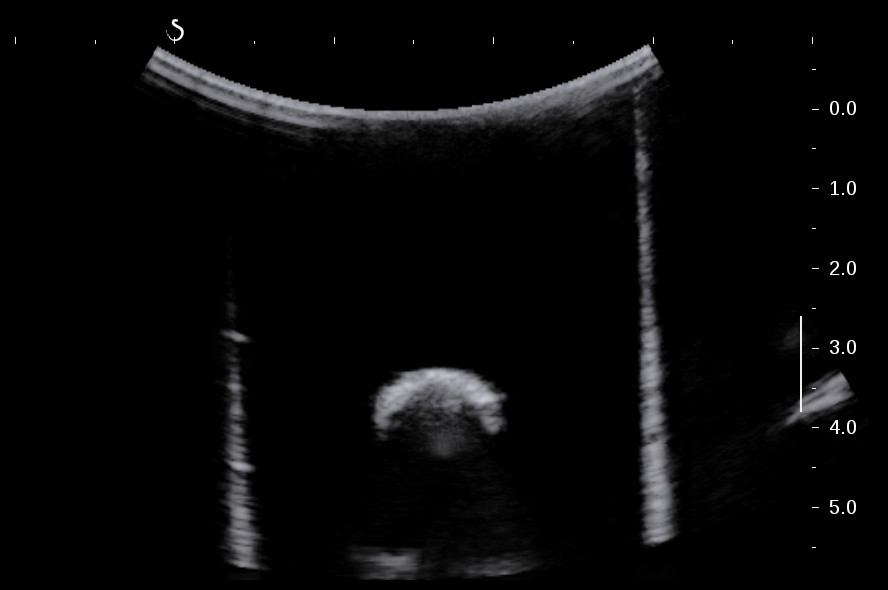}
\caption{Dataset 1 setup and corresponding B-scan. }
\label{fig:dataset1_imgs}
\end{figure}

\begin{figure}[!htb]
\centering
\includegraphics[width=12cm,height=7cm]{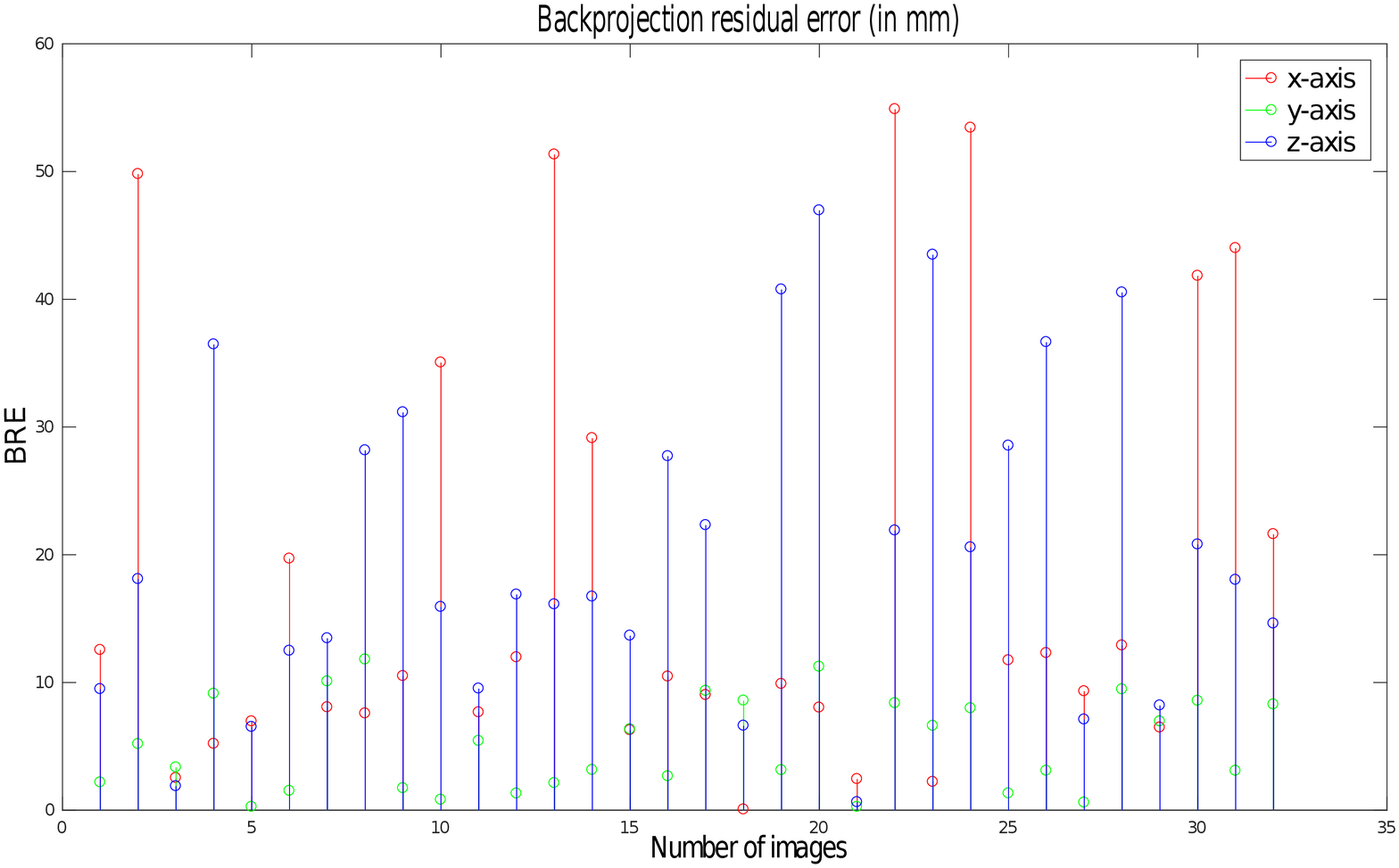}
\caption{Backprojection error for Dataset 1}
\label{fig:backdata1}
\end{figure}

In the first trial of the approach, we used two different sizes of the checkerboard markers (Figure~\ref{fig:dataset1_imgs}). The 
BRE plot of this acquisition is shown in Figure~\ref{fig:backdata1}. With this approach, the BRE is very high along all axes (Table~\ref{table:std_data1}). The mean error was greater than 5~$mm$. The error predominantly occurred due to incorrect pose estimations because of the different size of the two markers.

\begin{table}[!htb]
\centering
\caption{BRE along axes for Dataset 1}
\label{table:std_data1}
\begin{tabular}{|l|l|l|l|}
\hline
& x & y & z  \\ \hline
Std. & 16.7091  &	 3.5429& 12.2821\\ \hline
Mean& 17.9636 &	5.1218& 20.3735\\ \hline
Min & 0.0592&	0.2596&	0.6292\\ \hline
Max& 54.8793&	11.7972&	46.9600\\ \hline
\end{tabular}
\end{table}

\subsubsection{Dataset 2}
\begin{figure}[!htb]
\centering
\includegraphics[width=12cm,height=7cm]{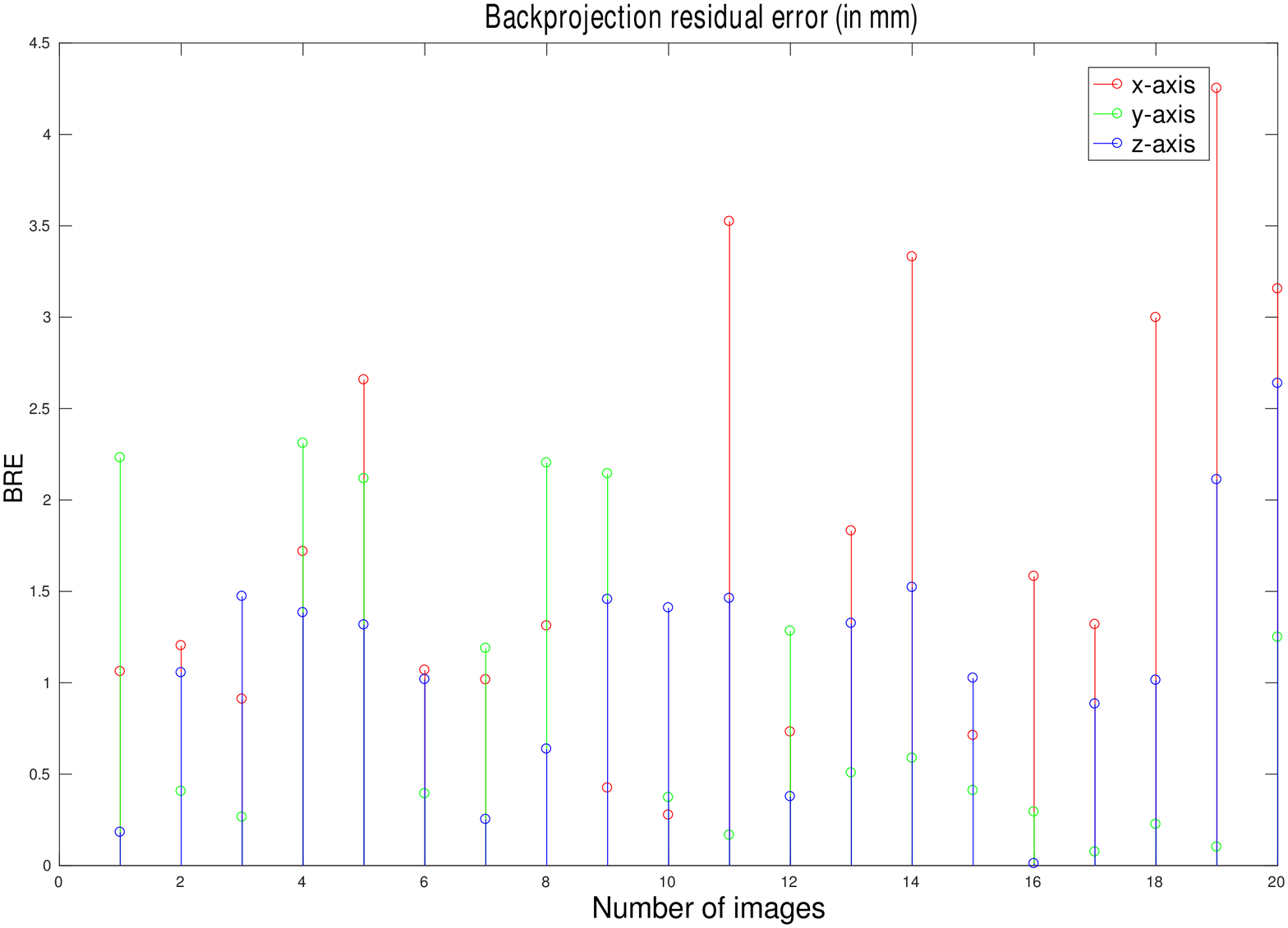}
\caption{Backprojection error for Dataset 2}
\label{fig:back2}
\end{figure}

The second dataset was acquired with the same size of both markers. We used 8 $\times$ 8 board with 16~$mm$ checker size. 
\begin{figure}[!htb]
\centering
\includegraphics[width=7cm,height=5cm]{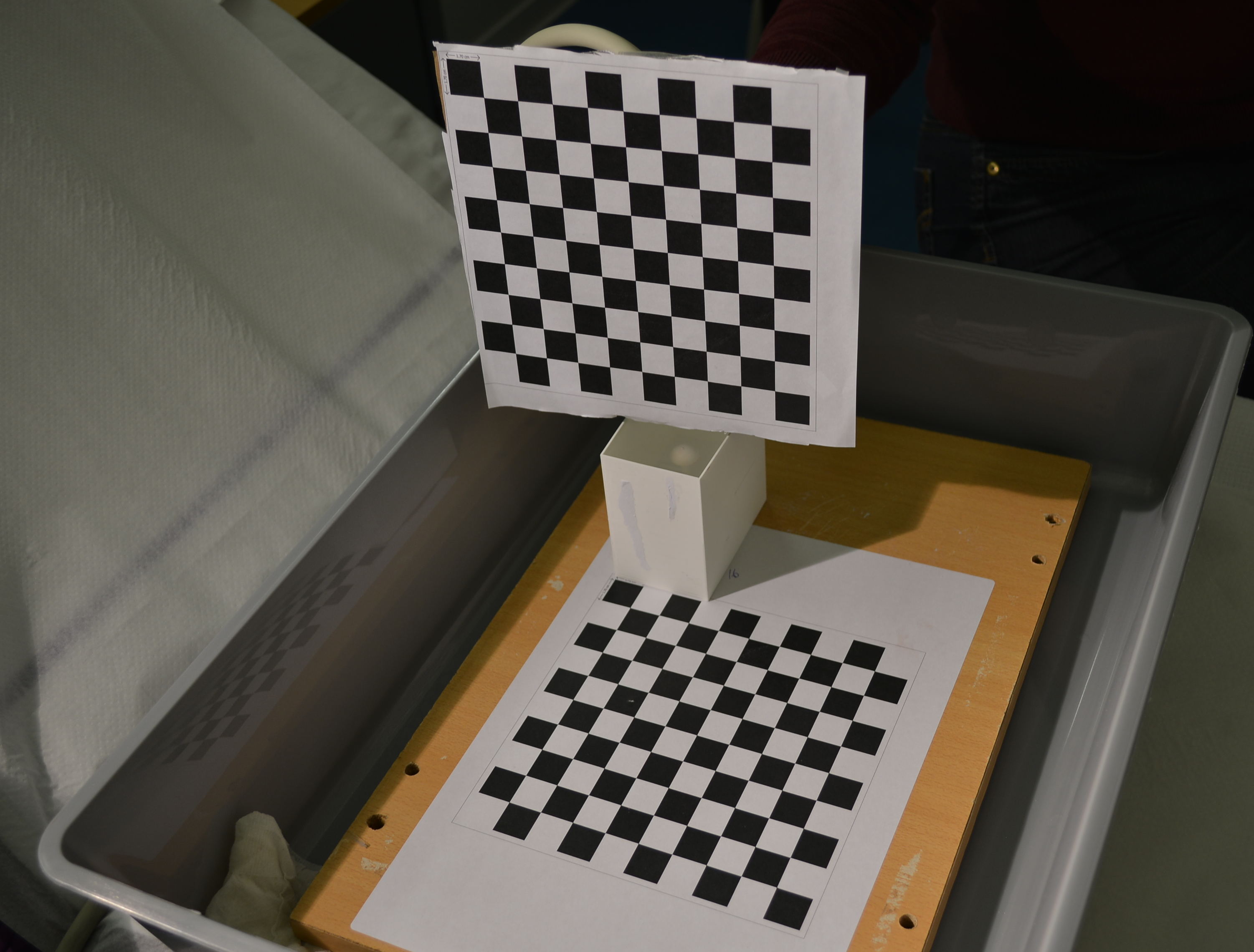}
\includegraphics[width=7cm,height=5cm]{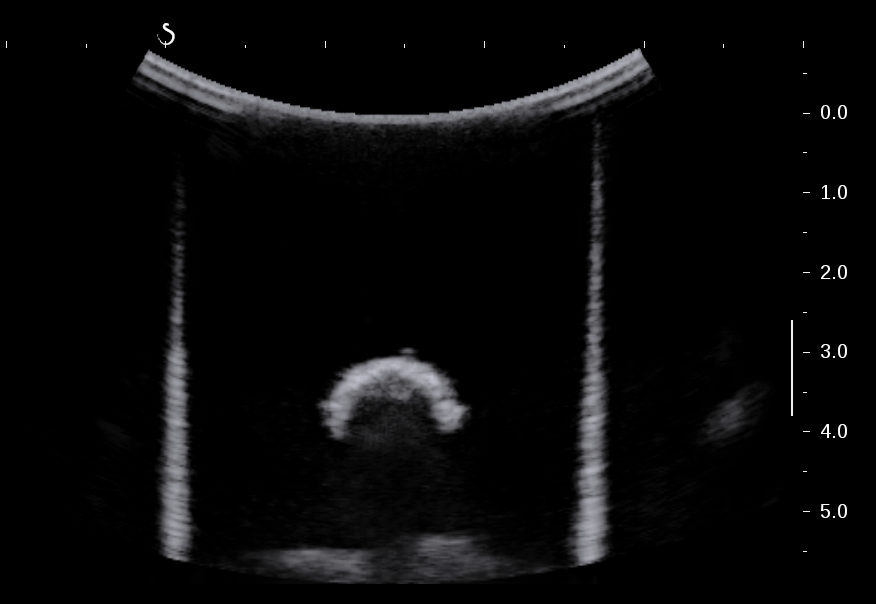}
\caption{Dataset 2 setup and corresponding B-scan. }
\label{fig:dataset2_imgs}
\end{figure}
This dataset contained 20 images used for the calibration. The results with this dataset are good compared to Dataset 1. We obtained near-millimeter accuracy for this dataset (Table~\ref{table:std_data2}). However, the maximum error is approximately  4~$mm$ along X-axis for few images 11, 14 16 (Figure~\ref{fig:back2}).

\begin{table}[!htb]
\centering
\caption{BRE along axes for Dataset 2}
\label{table:std_data2}
\begin{tabular}{|l|l|l|l|}
\hline
& x & y & z  \\ \hline
Std. & 1.1225  &	 0.8118& 0.6235\\ \hline
Mean& 1.754&	0.926& 1.127\\ \hline
Min & 0.276&	0.074&	0.011\\ \hline
Max& 4.252&	2.311&	2.638\\ \hline
\end{tabular}
\end{table}

%\begin{figure}[htb]
%	\centering
%	\epsfxsize=4cm
%	{\epsfbox{figures/backprojData3.eps}}
%\caption{Backprojection error for ArUco system}
%\label{fig:back}
%\end{figure}

%\newpage
\subsubsection{Dataset 3}
 This dataset was acquired with more number of checkers. We used 11 $\times$ 11 boards with 13$mm$ checker size. This dataset was captured with a set of 50 scans. 
 
 \begin{figure}[!htb]
 \centering
 \includegraphics[width=7cm,height=5cm]{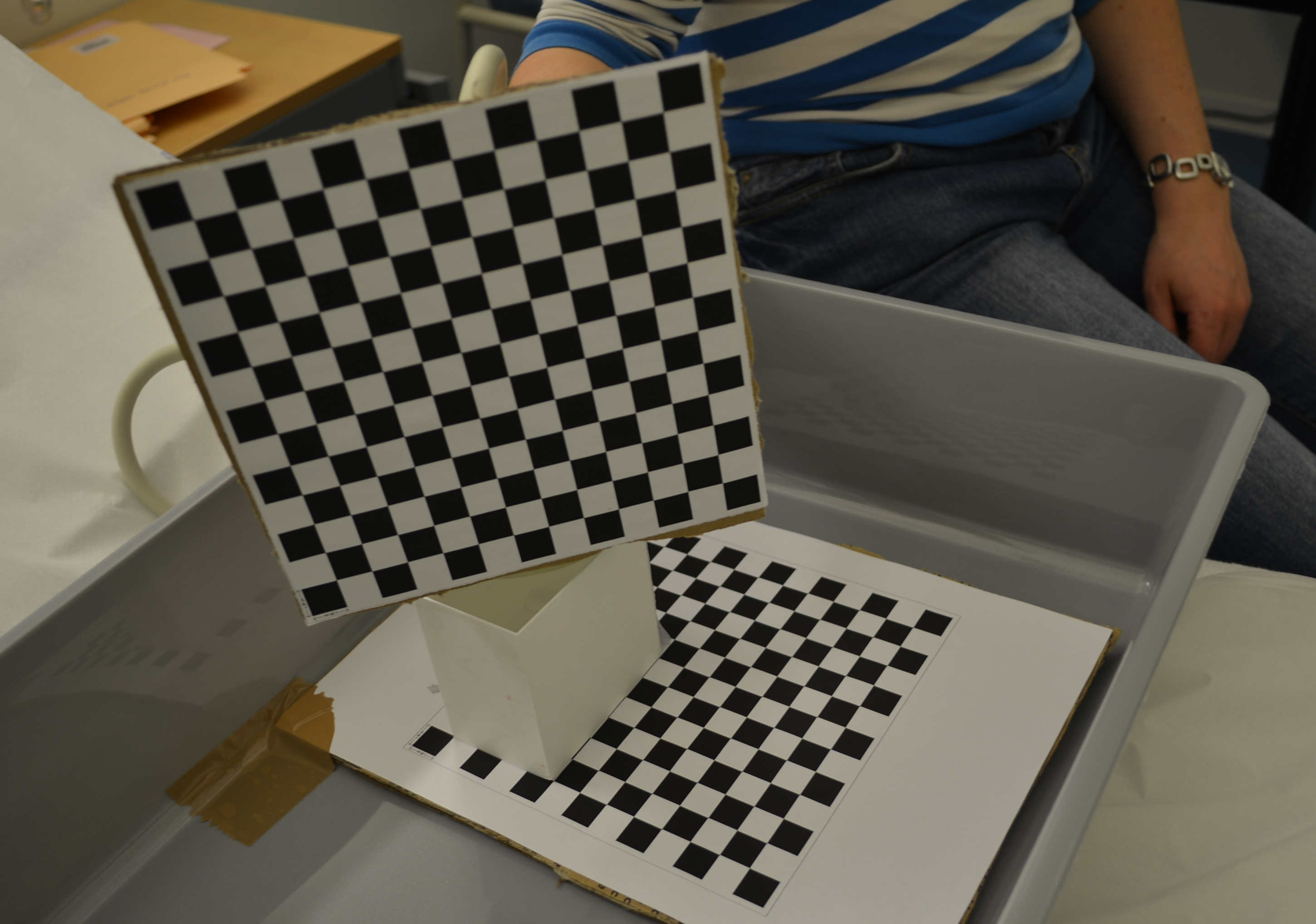}
 \includegraphics[width=7cm,height=5cm]{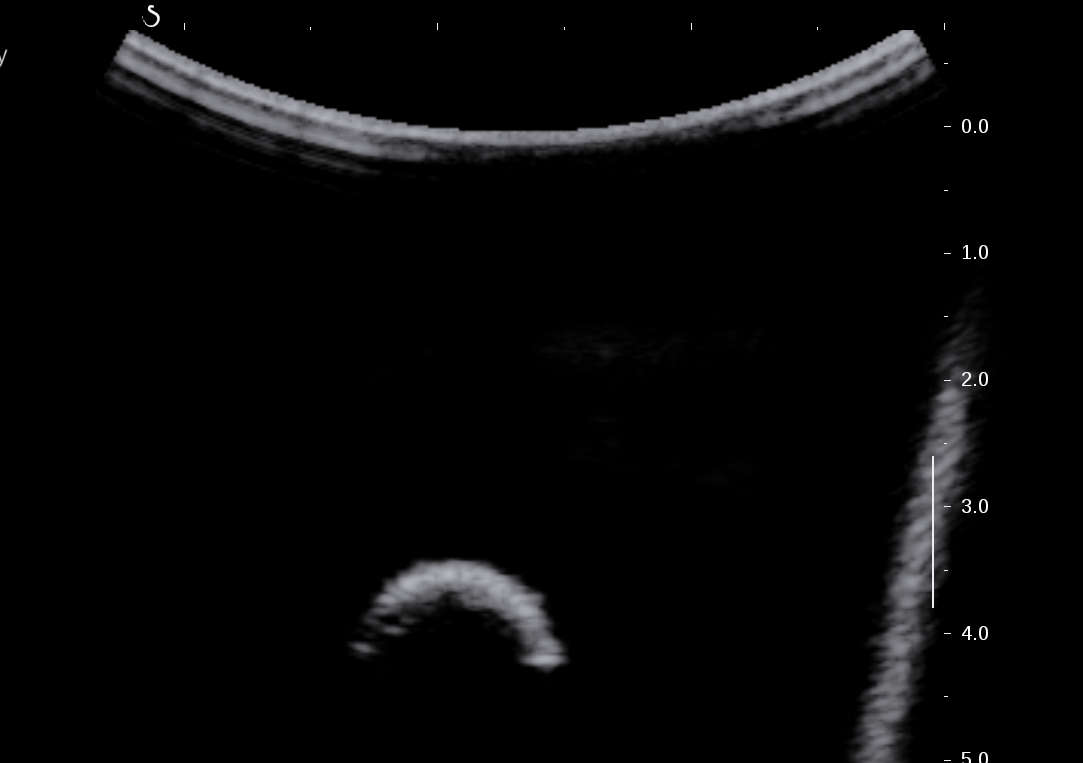}
 \caption{Dataset 3 setup and corresponding B-scan. }
 \label{fig:dataset3_imgs}
 \end{figure}
  
\begin{figure}[!htb]
\centering
\includegraphics[width=12cm,height=7cm]{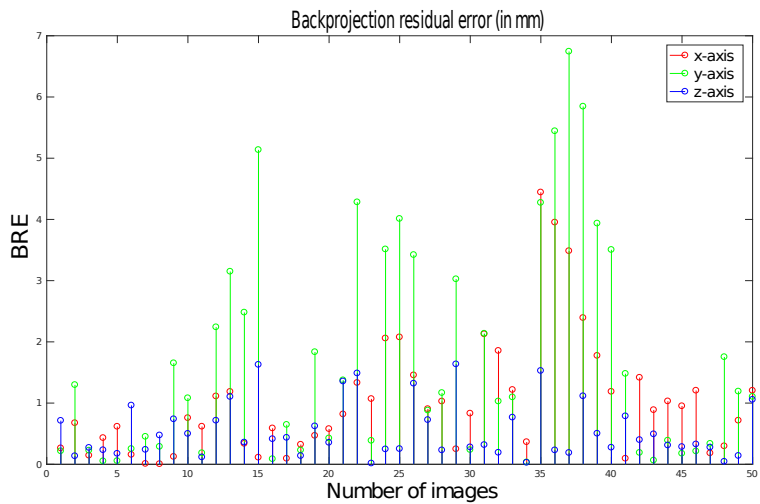}
\caption{Backprojection error for Dataset 3}
\label{fig:DataBackProjGoodCheck}
\end{figure}

 This dataset was evaluated with a different number of images. Table~\ref{table:std_data3} lists the values with different number of images. We first calculated error with all number of images i.e images with any error ($<\infty$). The maximum error with all images is 6~$mm$ along Y-axis. We can see that most of the contributed BRE (Figure~\ref{fig:DataBackProjGoodCheck}) is along the Y-axis. 
 The performance of this dataset was evaluated by removing high error images. 
 This dataset was evaluated by removing images with BRE $>2mm$, $>1.5mm$ and $>1mm$. The accuracy was improved after removing images with high BRE. A sub-millimeter accuracy was achieved in this acquisition. 

% Please add the following required packages to your document preamble:
% \usepackage{multirow}
% Please add the following required packages to your document preamble:
% \usepackage{multirow}
\begin{table}[!htb]
\centering
\caption{BRE along axes for Dataset 3}
\label{table:std_data3}
\begin{tabular}{|c|l|l|l|l|l|}
\hline
\multicolumn{1}{|l|}{Axis} & Parameter & \multicolumn{4}{c|}{Error in mm (Number of images)}   \\ \hline
\multirow{5}{*}{X}         &           & $<\infty$ (50) & $<2$ (32)  & $<1.5$ (25) & $<1$ (14) \\ \cline{2-6} 
                           & Std.      & 0.959     & 0.342 & 0.310  & 0.242 \\ 
                           & Mean      & 1.024     & 0.449 & 0.418  & 0.298 \\ 
                           & Min       & 0.005     & 0.020 & 0.047  & 0.027 \\ 
                           & Max       & 4.443     & 1.741 & 1.453  & 0.795 \\ \hline
\multirow{4}{*}{Y}         & Std.      & 1.758     & 0.570 & 0.281  & 0.139 \\ 
                           & Mean      & 1.704     & 0.608 & 0.392  & 0.161 \\  
                           & Min       & 0.034     & 0.008 & 0.012  & 0.003 \\ 
                           & Max       & 6.743     & 2.433 & 0.944  & 0.509 \\ \hline
\multirow{4}{*}{Z}         & Std.      & 0.444     & 0.285 & 0.249  & 0.202 \\ 
                           & Mean      & 0.542     & 0.350 & 0.296  & 0.256 \\  
                           & Min       & 0.016     & 0.008 & 0.019  & 0.066 \\ 
                           & Max       & 1.635     & 1.132 & 0.948  & 0.833 \\ \hline
\end{tabular}
\end{table}

\subsection{ArUco system}\label{sec:aruco_result}
\begin{figure}[!htb]
\centering
\includegraphics[width=12cm,height=7cm]{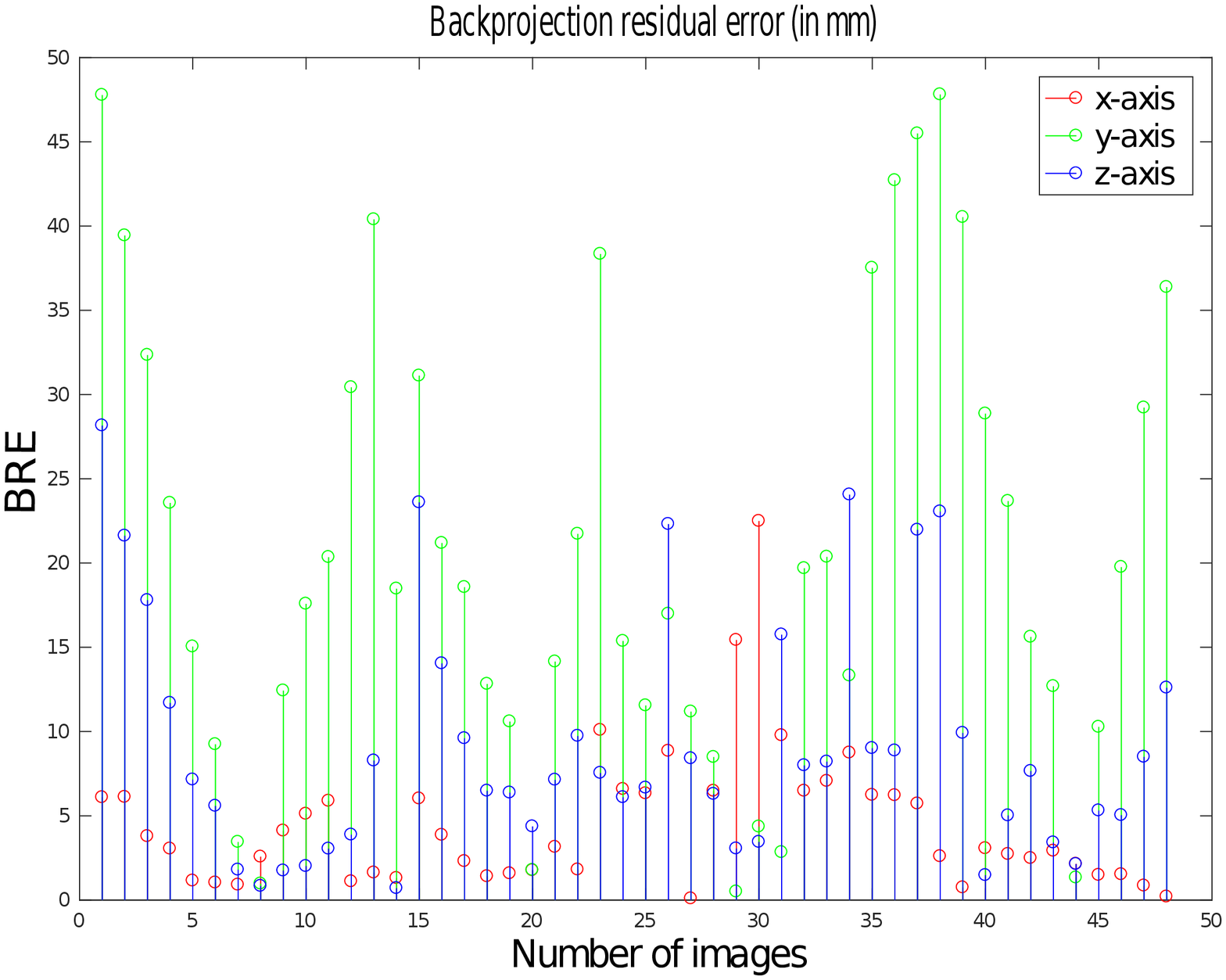}
\caption{Backprojection error for ArUco system}
\label{fig:ArucoBackProj}
\end{figure}

The system was also exploited with ArUco pose estimation. This approach was implemented with OpenCV + ArUco library. The results (Table~\ref{table:std_data_aruco}) and BRE (Figure~\ref{fig:ArucoBackProj}) with ArUco showed high error along all the axes. The highest error lies along the y-axis. We could not analyze the exact cause for  such high error in the pose estimation with ArUco. However, we hypothesize that  multiple causes could have contributed to this error which are: inaccurate measurements, camera movement during the acquisition, phantom movement and inaccurate camera calibration (high re-projection error, high distortion coefficients). 

\begin{figure}[!htb]
\centering
\includegraphics[width=7cm,height=5cm]{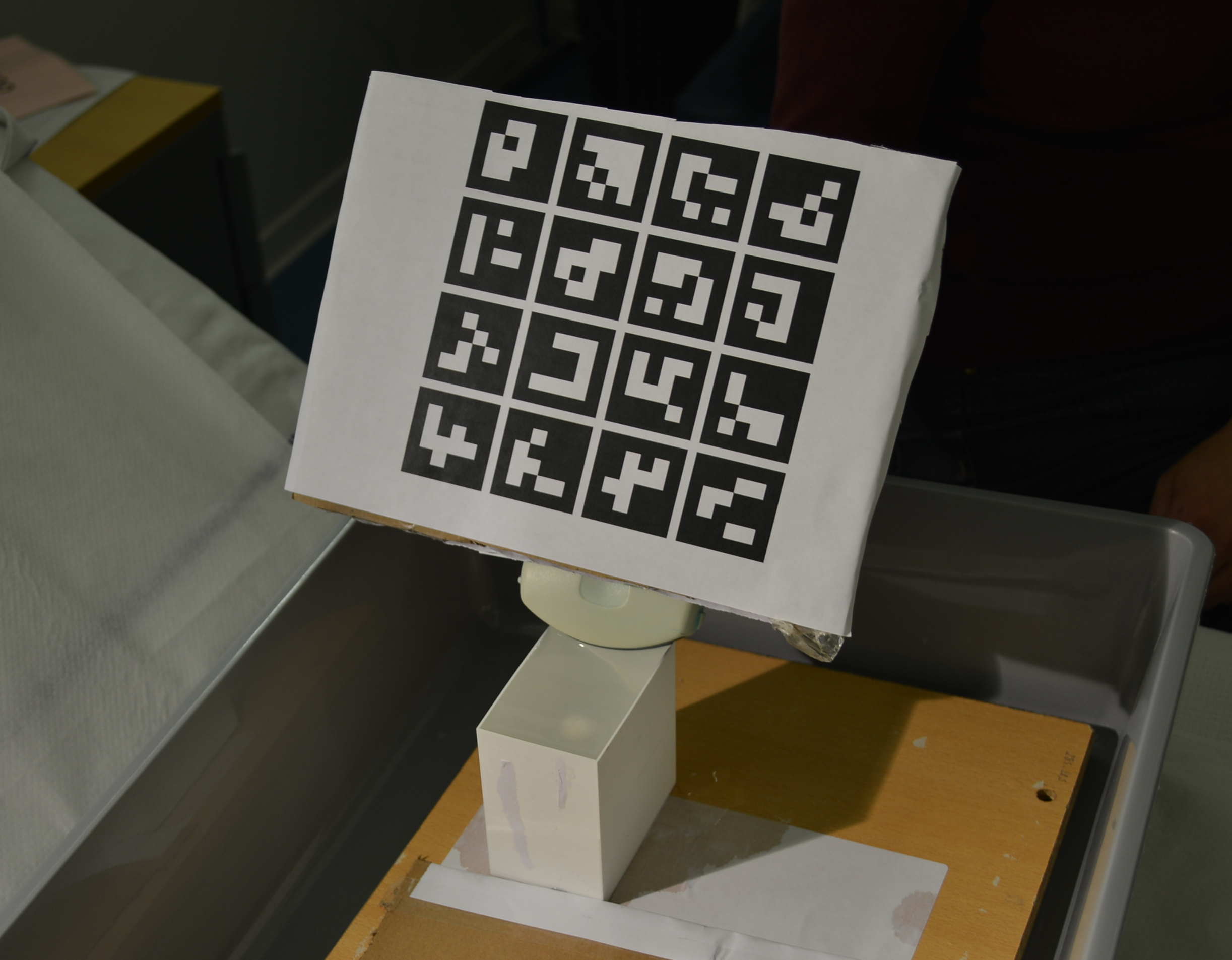}
\includegraphics[width=7cm,height=5cm]{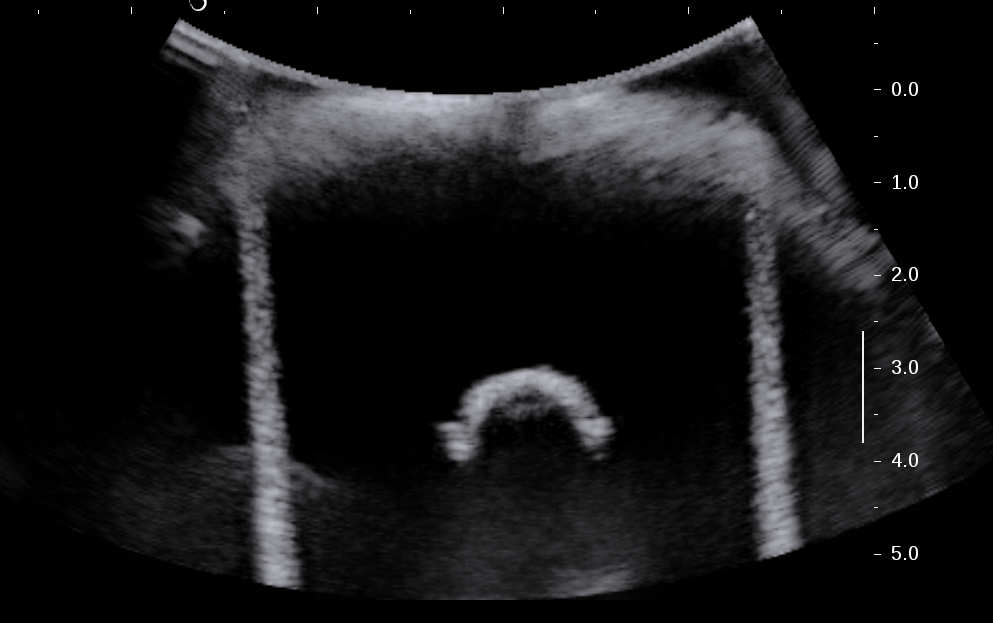}
\caption{Dataset ArUco setup and corresponding B-scan. }
\label{fig:datasetaruco_imgs}
\end{figure}

\begin{table}[!htb]
\centering
\caption{BRE along axes for Dataset ArUco}
\label{table:std_data_aruco}
\begin{tabular}{|l|l|l|l|}
\hline
 & x-axis  & y-axis & z-axis  \\ \hline
Std. & 4.0703 &  13.3303  &  7.0075\\ \hline
Mean& 4.4380 &  20.7976  &  9.1536\\ \hline
Min & 0.0908 &   0.4992  &  0.7023\\ \hline
Max& 22.4928 &  47.8121 &  28.1625\\ \hline
\end{tabular}
\end{table}

%\section{Conclusion}
%\section{Future Work}
%\section{Applications}
\chapter{Conclusion} \label{chap:Conclusion}

\section{Summary}
Freehand US imaging with conventional 2D transducers is useful in many applications including 3D US volume reconstruction, Image Registration with MRI/CT, needle guided therapies etc. To realize all these applications, the probe calibration is a preliminary step. While several calibration methods have been presented, there is no single method existing which outperforms all the others. In the proposed system, a simple, cost-effective and accurate calibration method was achieved. 

In this work, two different pose estimation approaches were studied and evaluated for the calibration task. The aim of the thesis to map the US image into the camera coordinate system has been achieved. Three different phantoms were designed and tested. We chose a point based hemisphere phantom to realize this task because of its simplicity and distinct features with no or few artifacts.  

In our approach using a checkerboard system we achieved sub-millimeter accuracy. The proposed system is cost-effective, simple and easy-to-use compared to existing expensive and complex systems. The errors in this system are considered reasonable compared with the existing systems. However, the overall accuracy of the system depends on the pose estimation accuracy, phantom construction and data acquisition. This accuracy is predominantly impeded by the pose estimation accuracy which depends on different factors such as camera calibration parameters, lighting conditions, clear line of sight and rigidness of the camera during acquisition. The AR marker based system resulted in high error in the calibration process. This error could not be analyzed with a single dataset and should be further investigated to make an accurate and robust system.

\section{Future Work}
Although the proposed system provides sub-millimeter accuracy, it should be further validated with more experiments and datasets. The reproducibility of the proposed system should be further validated with more datasets. Calibration reproducibility (CR) should be measured to validate the performance on new set of images as proposed in \cite{lindseth2003probe}. With the current system, the high error images are removed by hard thresholding to improve the accuracy. In the future, we propose to remove such outliers or high error data. This can be achieved with widely used RANSAC or least median squares (LMeDS) algorithms.

The accuracy of the checkerboard system depends on its size. It is difficult to have the big size markers in the clinical environment as used within this study. This problem can be solved by using Augmented Reality (AR) markers like ArUco. Although we have used a plane of ArUco markers, we can use small sized markers to realize the system. A large error was estimated with ArUco system and we could not validate it with more experiments. However, more accurate and robust system can be achieved using ArUco markers with correct and reliable data acquisition. In addition, the phantom can be further redesigned with multiple spheres to improve the precision. This will also enhance the robustness and accuracy of the system. 
Although the proposed system has been designed for the application to image fusion with MRI/CT, it can be modified to reconstruct the 3D US volume in real time. The future work includes improving accuracy with ArUco and develop the system for the application of image fusion with US-MRI/CT. 
%\section{Discussion}

%\section{Applications}
%\appendix
%\input{appendix.tex}

%   this is for BibTeX.  remove if you plan to write the references in the document
\bibliographystyle{plain}
\bibliography{refs}

%adds the bibliography to the table of contents
\addcontentsline{toc}{chapter}
         {\protect\numberline{Bibliography\hspace{-96pt}}}

\end{document}